\documentclass[preprint,10pt]{elsarticle}
\usepackage{lineno,hyperref}
\modulolinenumbers[5]
\journal{Computers \& Security (Elsevier) on 14, June 2020  and  accepted on 19, October 2020.}

\usepackage{makecell}
\usepackage{graphics}
\usepackage{amsmath}
\usepackage{adjustbox}
\usepackage{color}
\usepackage{lineno}
\usepackage[margin=1in]{geometry}
\usepackage{setspace} 

\newcommand{\EN}[1]{\textcolor{blue}{{#1}}}

\begin{document}
\begin{frontmatter}

\title{A Survey of Machine Learning Techniques in Adversarial Image Forensics}

\author[rvt]{Ehsan Nowroozi}
\ead{ehsan.nowroozi@diism.unisi.it}
\author[focal]{Ali Dehghantanha}
\ead{adehghan@uoguelph.ca}
\author[rvt1]{Reza M. Parizi}
\ead{rparizi1@kennesaw.edu}
\author[els2]{Kim-Kwang Raymond Choo}
\ead{raymond.choo@fulbrightmail.org}

\address[rvt]{Department of Information Engineering and Mathematics, University of Siena, Siena, Italy}
\address[focal]{School of Computer Science, University of Guelph, Ontario, Canada}
\address[rvt1]{College of Computing and Software Engineering, Kennesaw State University, GA, USA}
\address[els2]{Department of Information Systems and Cyber Security, University of Texas at San Antonio, San Antonio, Texas, USA}

\begin{abstract}
Image forensic plays a crucial role in both criminal investigations (e.g., dissemination of fake images to spread racial hate or false narratives about specific ethnicity groups) and civil litigation (e.g., defamation). Increasingly, machine learning approaches are also utilized in image forensics. However, there are also a number of limitations and vulnerabilities associated with machine learning-based approaches, for example how to detect adversarial (image) examples, with real-world consequences (e.g., inadmissible evidence, or wrongful conviction). Therefore, with a focus on image forensics, this paper surveys techniques that can be used to enhance the robustness of machine learning-based binary manipulation detectors in various adversarial scenarios. \\

 
\end{abstract}

\end{frontmatter}

\section{Introduction}
\label{sec.intro}
As consumer technologies (e.g., image acquisition and editing tools) and artificial intelligence techniques advance, editing digital images and creating fake images are becoming easier and cheaper. Deliberate manipulation of digital images can be innocuous (e.g., to improve the quality and appearance of an image) or carried with malicious intent (e.g., to alter the semantic content of the image, or to establish an alibi). The diffusion of fake images has implications on judicial systems, global economy, financial health, and even homeland and national security. Not surprisingly, there have been interest from the digital forensics, and more specifically image forensics, community in recent years to detect deliberate manipulation of digital images. There have also been interest from the commercial market, as suggested in a recent study \cite{statistic}.

Image forensics, an emerging forensic discipline, seeks to determine the history of an image (e.g., its origin), the processing it underwent, etc, in order to determine the authenticity of the images \cite{Piva2013SurveyForensics}. In other words, key image forensic tasks include \textit{source classification} (to reveal the source image produced from an image acquisition device, such as scanner and camera), \textit{source identification} (to identify the acquisition device used to take the image), \textit{reverse engineering of processing operators} (to identify the chain of processing operators, and processing includes deliberate manipulation), and \textit{authenticity verification} (to determine whether the image has been manipulated).

In recent years, there have also been attempts to utilize machine learning (ML)-based techniques to both support different image forensics tasks and to defeat ML-based image forensics \cite{carlini2020evading,goodfellow2014explaining,guarnera2020deepfake,li2020celeb}. This reinforces the importance of developing techniques to protecting ML systems (also referred to as adversarial ML (Adv-ML)). Both ML and deep learning algorithms can be vulnerable to adversarial attacks that hinder their applications in security-sensitive domains, such as image forensics. For example, a deep neural network may report high confidence in a wrong prediction or can be circumvented by image perturbation techniques. Therefore, there have been attempts to design effective adversarial counter forensics (CF) techniques. 

A number of literature surveys and reviews on the applications of ML-techniques in image forensics has been published in the literature \cite{Dario2016,Nicolas2016,Akhtar2018VisionSurvey,Xue2020MLsec} and in image forensics \cite{FERREIRA2020,Yang2020, Kaur2020, verdoliva2020media}, although adversarial image forensics is generally not discussed. Amodei et al., \cite{Dario2016}, for example, reviewed the general security concerns in artificial intelligence, particularly reinforcement learning and supervised learning algorithms.   
A general review of security implications on the use of ML approaches and their countermeasures was presented in \cite{Nicolas2016,Xue2020MLsec}. Akhtar et al.  \cite{Akhtar2018VisionSurvey} focused on adversarial attacks on deep learning approaches in computer vision. However, there have been limited studies focusing on ML-security issues in (adversarial) image forensics, a gap we seek to address in this paper.

\begin{figure}[h!]
		\centering
		\includegraphics[width=0.99\columnwidth]{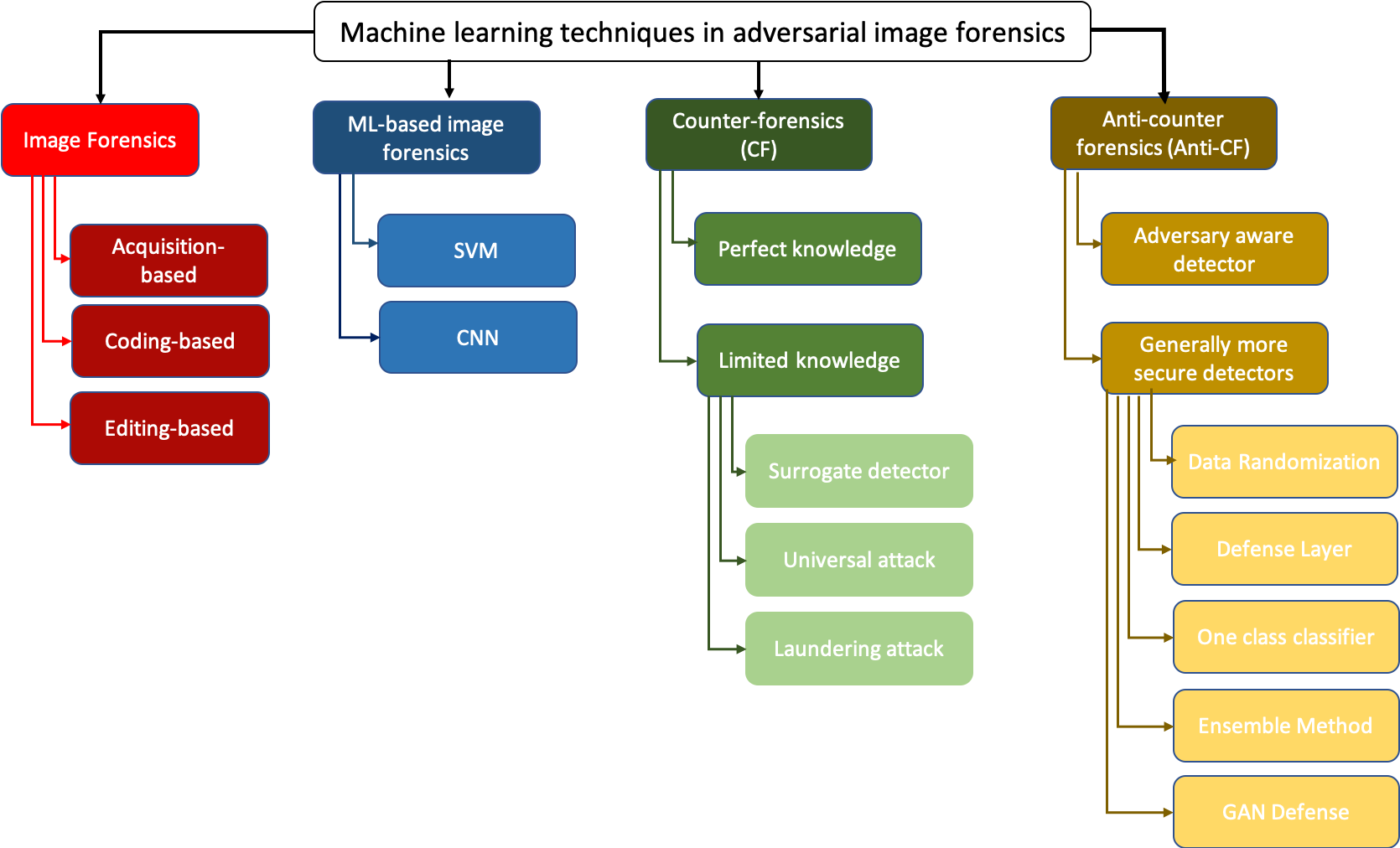}
		\caption{Graphical abstract of adversarial image forensics}
		\label{Graph_ABS.Fig}		
\end{figure}

Specifically, in this paper we survey existing ML techniques for image forensics, including those that can be utilized in the adversarial setting (e.g., image manipulation), and CF. In the survey, we also reviewed the various approaches that can be used to enhance the security of binary manipulation detectors based on ML and defensive techniques during the testing stage. Figure \ref{Graph_ABS.Fig} shows a graphical abstract concerning the application of machine learning techniques in adversarial image forensics.

We will now explain the layout of the remaining of this paper. Section \ref{section2} introduces the reader to relevant background materials. In Section \ref{section3}, we present the survey on ML-based image forensics, prior to reviewing adversarial image forensics in Section \ref{section4}. In the latter section, we also revisit ML-based CF challenges and potential mitigation approaches (commonly referred to as anti-CF). Moreover, a classification of ML techniques in the presence of the adversary is presented. In Section \ref{sec.conc}, we conclude this paper.


\section{Background} \label{section2}

Here, we will revisit popular image forensics approaches such as Support-Vector- Machines (SVM) and Convolutional-Neural-Networks (CNNs), followed by ML-based image forensic techniques.


\subsection{Image Forensic Analysis Approaches}
\label{Methods_Forensics}

The primary goal of image forensics is to identify remnants of any activities over an image file, which can be utilized by a forensic analyst in timeline or relational analysis and event reconstruction. Existing image forensic analysis techniques can be broadly categorized into those based on \textit{image acquisition}, \textit{image coding}, and \textit{image processing} \cite{Piva2013SurveyForensics}. 
	
\begin{itemize}
	\item \textbf{Remnants of Acquisition-based Activities} Every step of the image acquisition process leaves some traces within an image (e.g., traces for instance of the particular Color Filter Array (CFA) pattern and the type of filter used for color interpolation) \cite{Piva2012CFA}. Thus, the digital image produces evidence for interpolation filter and CFA pattern \cite{Khanna2007ScannerIU}. Furthermore, a camera sensor, by default, leaves a particular noise in each captured image, which is known as Photo-Response-Non-Uniformity (PRNU). In other words, two different image acquisition devices of the same make and model are likely to leave different patterns and the analysis of the traces will allow us to detect the image acquisition device used to acquire the image. 
	Khanna et al. \cite{Khanna2007ScannerIU} detect footprints of digital tampering in lossless and lossy in compressed images. Table \ref{Forensic-Analysis_AC.tab} summarizes existing forensic analysis methods designed to identify remnants of image acquisition activities.
\end{itemize}

\begin{itemize}
	\item \textbf{Remnants of Coding-based Activities} Modern-day digital cameras generally utilize JPEG compression as a standard format for efficient storage and transmission, which partly explains the interest in the study of an image's compression history \cite{Luo2010JPEG,Bianchi2011JPEG,Ehsan2015JPEG}. Since various imaging software applications generally consider different compression parameters and employ different quantization tables \cite{Luo2010JPEG}, the analysis of inconsistencies in quantization matrices can also be used for source and forgery detection. 
	Ferrara et al. \cite{Piva2012CFA} proposed a method to distinguish aligned from not-aligned DJPEG compression by identifying the DCT coefficients of the first compression. Table \ref{Forensic-Analysis_Code.tab} summarizes existing forensic analysis methods designed to identify remnants of image coding activities.
\end{itemize}

\begin{itemize}
	\item \textbf{Remnants of Editing-based Activities} 
	The use of image and video editing software will also result in the presence of digital footprints. For example,the application of a geometric transformation, such as a rotation or a resizing, requires interpolation of pixel values that leaves detectable traces in the image \cite{Piva2013SurveyForensics}. By employing a resizing and rotation geometric transformation, the software leaves detectable artifacts on the image due to value interpolation. Moreover, by altering the subpart of the image in terms of applying blurriness, contrast, saturation, etc, the software leaves artifacts in different areas of the image \cite{stamm2010forensic}. Median filtering in image processing, plus denoising and smoothing, can additionally be used to hide footprints of previous processing attempts \cite{Yuan2011MedianFilter}.  Yuan et al. \cite{Yuan2011MedianFilter}, for example, proposed a novel techniques for detecting median filter images through block-wise method by considering median filter between the blocks.
\end{itemize}

\begin{table*}[h!]
	\renewcommand\arraystretch{3.0}
	\centering
	\begin{center}
		\caption{Acquisition-based forensic analysis methods: A comparative summary}
		\resizebox{\columnwidth}{!}{%
			\begin{tabular}{| c | c |}
				\hline
				 \bf{Reference} & \bf{Results}  \\
				\hline
				 \cite{Piva2012CFA} &  \makecell{\\Results on the median value of various datasets and interpolations, \\ \textbf{No CFA:} bilinear 1.58, bicubic 1.55, gradient 1.67, median 1.81 \\ \textbf{Ideal:} bilinear 1.16, bicubic 2.13, gradient 2.04, median 2.01 \\ \textbf{Canon EOS:} bilinear 2.00, bicubic 1.90, gradient 1.89, median 1.96 \\ \textbf{Nikon D50:} bilinear 1.73, bicubic 1.79, gradient 1.83, median 1.81 \\ \textbf{Nikon D7000:} bilinear 2.20, bicubic 2.06, gradient 1.72, median 1.89 \\ \textbf{Nikon D90:} bilinear 1.99, bicubic 1.92, gradient 1.66, median 1.92. \\}  \\
				\hline
				\cite{Khanna2007ScannerIU} & \makecell{\\Accuracy are achieved based on four different scanners: ($S_1$) Epson Perfection 4490 Photo, \\($S_2$) HP ScanJet 6300c-1, ($S_3$) HP ScanJet 6300c-2, and ($S_4$) HP ScanJet 8250.  \\
				\textbf{2D Reference Pattern,} \\
				The accuracy of actual $S_1$ is predicted with 66.8\% as $S_1$ and 33.2\% as a $S_2$. \\
				The accuracy of actual $S_2$ is predicted with 22.5\% as $S_1$ and 77.5\% as a $S_2$. \\
				The accuracy of actual $S_2$ is predicted with 69.4\% as $S_2$ and 30.6\% as a $S_4$. \\
				The accuracy of actual $S_4$ is predicted with 0.40\% as $S_2$ and 99.6\% as a $S_4$. \\
			    \textbf{1D Reference Pattern,} \\   
			    The accuracy of actual $S_1$ is predicted with 63.7\% as $S_1$ and 36.3\% as a $S_2$. \\ 
			    The accuracy of actual $S_2$ is predicted with 21.6\% as $S_1$ and 78.4\% as a $S_2$. \\
			    The accuracy of actual $S_2$ is predicted with 85.1\% as $S_2$ and 14.9\% as a $S_4$. \\
			    The accuracy of actual $S_4$ is predicted with 0.00\% as $S_2$ and 100.0\% as a $S_4$. \\
		    \\ }\\
				\hline

		\end{tabular}
			\label{Forensic-Analysis_AC.tab}
		}
		
	\end{center}
	
\end{table*}
\begin{table*}[h!]
	\renewcommand\arraystretch{3.0}
	\centering
	\begin{center}
		\caption{Coding-based of forensic analysis methods: A comparative summary}
		\resizebox{\columnwidth}{!}{%
			\begin{tabular}{| c | c |}
				\hline
				 \bf{Reference} & \bf{Results}  \\
				\hline
				 \cite{Luo2010JPEG} &  \makecell{\\ \textbf{QF 98}, the accuracy for image size $256 \times 256$ is 92.36\%, $128 \times 128$ is 93.65\%, \\ $64 \times 64$ is 93.94\%, $32 \times 32$ is 92.91\%, $16 \times 16$ is 90.32\%, and $8 \times 8$ is 81.95\%. \\		    
			    \textbf{QF 85}, the accuracy for image size $256 \times 256$ is 92.95\%, $128 \times 128$ is 94.21\%, \\ $64 \times 64$ is 94.44\%, $32 \times 32$ is 93.71\%, $16 \times 16$ is 92.68\%, and $8 \times 8$ is 89.06\%. \\		    
			    \textbf{QF 75}, the accuracy for image size $256 \times 256$ is 92.95\%, $128 \times 128$ is 94.21\%, \\ $64 \times 64$ is 94.42\%, $32 \times 32$ is 93.69\%, $16 \times 16$ is 92.68\%, and $8 \times 8$ is 88.75\%. \\}\\
				\hline
				 \cite{Bianchi2011JPEG} & \makecell{\\ $QF_1$ 50 and $QF_2$ 50, 60, 70, 80, 90, and 100, AUCs are 0.50, 0.90, 1.00, 1.00, \\0.99, and 0.99 respectively. \\ $QF_1$ 70 and $QF_2$ 50, 60, 70, 80, 90, and 100, AUCs are 0.77, 0.83, 0.49, 1.00, \\0.99, and 0.99. \\ $QF_1$ 90 and $QF_2$ 50, 60, 70, 80, 90, and 100, AUCs are 0.58, 0.63, 0.70, 0.78, \\0.50, and 0.99. \\}\\
				\hline

			\end{tabular}
			\label{Forensic-Analysis_Code.tab}
		}
		
	\end{center}
	
\end{table*}
\begin{table*}[h!]
	\renewcommand\arraystretch{3.0}
	\centering
	\begin{center}
		\caption{Editing-based forensic analysis methods: A comparative summary}
		\resizebox{\columnwidth}{!}{%
			\begin{tabular}{| c | c | c |}
				\hline
				\bf{Category}  & \bf{Reference} & \bf{Results}  \\
				\hline
				\makecell{contrast \\enhancement} & \cite{stamm2010forensic} & \makecell {\\ In this scenario, the detection algorithm's performance increases with high parameter $c$. \\ 
				They achieved best performance when $c=112$. The performance degrades \\when they consider $c=32$. This scenario applies to contrast enhancement detection \\with $\gamma = 1.1$.\\} \\
				\hline
				\makecell{median \\ filtering} & \cite{Yuan2011MedianFilter} &  \makecell{\\ \textbf{No-JPEG and JPEG Images:} ROC curves for MMF and $f$ in \textbf{uncompressed} \\scenario demonstrates that features are more robust to detect median filtering. In \textbf{JPEG} scenario \\they consider $3 \times 3$ blocks and $QF$ = [100, 90, 80, 70]. The performance degrades \\ with $QF >$ 80 but the performance improves when $QF <$ 70.\\}\\
				\hline

			\end{tabular}
			\label{Forensic-Analysis_Editing.tab}
		}
		
	\end{center}
	
\end{table*}
%
\subsection{Image Forensics Tools}

Now, we will review three widely used image analysis tools (see also Table \ref{TOOLS}).

\begin{table*}[h!]
	\renewcommand\arraystretch{3.0}
	\begin{center}
		\caption{Image forensics tools: A comparative summary}
		\resizebox{\columnwidth}{!}{%
			\begin{tabular}{| c | c | c |}
				\hline
				\bf{Tools}  & \bf{Pros} & \bf{Cons} \\
				\hline
				\bf{Forensically}  & \makecell{Detect clone detection, metadata extraction, \\magnifier, error level analysis, noise analysis.} & \makecell{Localization detection} \\
				\hline
				\bf{Assembler}  & \makecell{Image enhancement and splicing} & \makecell{detect only a few processings.} \\
				\hline
				\bf{JPEGsnoop}  & \makecell{extract the hidden details of the \\compressed image and motion JPEG.} & \makecell{Detect only one specific compression method \\e.g. JPEG not TIFF.} \\
				\hline
			\end{tabular}
			\label{TOOLS}
		}	
	\end{center}	
\end{table*}

\subsubsection{Forensically}

Forensically is a free tool in digital image forensics that offers features such as clone detection and metadata extraction \cite{Jonas_Wagner}. The tool's \textit{magnifier} or zoom factor helps analysts to find hidden features in an image by magnifying the pixels' size and color. Magnifier for detection uses the following three items based on \textit{histogram equalization}, \textit{auto contrast}, and \textit{auto contrast by channel}. \textit{Clone detector} in Forensically identifies manipulated areas of two or more image files. Similar areas are illustrated with blue and connected with red and overlapping areas with white color. Several options can be selected in this tool based on minimal similarity, minimal detail, minimal cluster size, block size, maximal image size, and show a quantized image. \textit{Error level analysis} compares the pristine image with a compressed version to identify possible features lost during compression. Therefore, manipulated areas can stand out in various ways. This tool compares different features of image files such as color quality, error scale, magnifier enhancement, and opacity. \textit{Noise analysis} is another part of Forensically, which is utilized as a median filter to identify noises in an image file that can be used for manipulation detection like airbrushing and wrapping. \textit{Level sweep} in Forensically helps analysts detect copy-pasting areas by considering sweep through an image histogram. Furthermore, the copy-pasting areas are more visible due to the use of magnifiers. With the \textit{JPEG analysis} tool, analyzers can extract metadata from JPEG files, such as quantization tables.

	
\subsubsection{Assembler}
Recently on 5 January 2020, Google published Assembler which is a tool designed for journalists to detect forged images \cite{Google}. Assembler uses several existing methods to recognize common image manipulation detection (e.g., image enhancement, copy-move, and splicing). This tool also includes a detector to identify deepfakes, which is generated through StyleGAN. Assembler can help one identify which part of the image has been manipulated, determining copy-move and splicing in the presence of image brightness. While Assembler can help journalists spot manipulated images, it does not include other existing manipulation methods for detecting video and audio files. Moreover, this system needs to be real-time when used. 

\subsubsection{JPEGsnoop}
JPEGsnoop \cite{JPEGsnoop} is a free software that can assess and extract hidden details of a compressed image, motion JPEG, and Photoshop files.  JPEGsnoop is also capable of analyzing the image's source to obtain information from compressed images such as quality factor, and reporting on information such as chrominance and luminance quantization matrix, estimation of JPEG quality factors, Huffman tables, and histogram for RGB images.


\subsection{ML-based Image Forensics}	
Unlike traditional image forensic tools, such as those discussed in Section \ref{Methods_Forensics}, ML algorithms can be used to learn complex patterns from a set of hand-crafted features and facilitate classification. 

\begin{itemize}
	\item \textbf{SVM-based image forensics} SVM is widely used, partly due to its simplicity and accuracy outcomes in many classification tasks. In most of the earlier SVM-based forensic approaches, the features are hand-crafted, extracted from the image based on some heuristics, and very specific for the problem at hand (e.g., to detect double JPEG compression (DJPEG) and recompression). For example, Chen et al. \cite{chen2008machine} considered a set of features to improve DJPEG traces to discriminate between double and single JPEG images, which can then be used to train an SVM-based classifier. 
	Nowroozi et al. \cite{Ehsan2015JPEG} considered the traces left by DJPEG image in the mean, variance, and entropy, by training the SVM classifier using these collective statistical features. Milani et al. \cite{Milani2012DiscJPEG} presented a statistic derived from the DCT histograms based on first significant digits (FSD), in order to distinguish single from double JPEG compression. 
	
Islam et al. \cite{Karmakar2018} introduced a technique for splicing and copy-move detection that utilizes SVM classification of Local Binary Patterns (LBP) descriptors derived from the block DCT of chroma channels. SVMs have  often been used for camera model identification, to fulfill multi-class classification based on high order features derived from the images or from the PRNU noise pattern generated by different cameras \cite{Filler2008PatternNoise}. A summary of the different SVM-based image forensics techniques is shown in Table \ref{SVM_Forensics.tab}.
	
	\begin{table*}[h!]
		\renewcommand\arraystretch{3.0}
		\centering
		\begin{center}
			\caption{Existing SVM-based image forensics techniques: A comparative summary}
			\resizebox{\columnwidth}{!}{%
				\begin{tabular}{| c | c | c |}
					\hline
					\bf{Reference} & \bf{Pros} & \bf{Cons}  \\
					\hline
					\cite{chen2008machine} & \makecell{Double JPEG detection.} & \makecell{1) Some cases failed (N/D). \\ 2) Performance degrades when SD is applied.} \\
					\hline
					\cite{Milani2012DiscJPEG} & \makecell{Distinguish between SJPEG and DJPEG.} & \makecell{Other CF strategies could deceive \\ this approach.} \\
					\hline
					\cite{Karmakar2018} & \makecell{Splicing and copy-move detection.} & \makecell{AUC decreases when window size\\ is $4 \times 4$.} \\
					\hline

				\end{tabular}
				\label{SVM_Forensics.tab}
			}
			
		\end{center}
		
	\end{table*}
\end{itemize}
\begin{itemize}
	\item \textbf{CNN-based image forensics} Deep learing (DL) methods such as Convolutional Neural Networks (CNNs) have also been used in steganalysis purposes, and generally the performance of these new CNN-based techniques exceeds those of classical model-based and standard ML-based techniques considerably.

The first task in image forensics utilizing CNNs for detecting median filtering is to apply feature representations with high accuracy compared with hand-crafted features methods  \cite{ChenMFwithCNN}. The authors in \cite{StammCNNuniv} studied a binary and a multi-class CNN, which is effective for the different manipulation operations such as blurring, median filtering, and resizing. This approach was later extended in \cite{BayarCNNUnivTIFS}. 
In each image patch, they found a strong difference between different camera models. However, these networks are partly shallow, consisting of only three or four convolutional layers. Besides, a pre-processing filtering step is employed to the first layer to force the network to look for the traces in the residual domain high-pass image, hence facilitating its job.

Niu et al. \cite{niu2019primary} computed a primary quantization matrix using DCT coefficients based on CNNs, which can work under a variety of situations. Also, they achieved good performance on small image patches. In summary, the method of considering completely self-learned features, without forcing the initial layers, has been shown to be useful, as long as sufficient training data is available. A summary of the different CNN-based image forensics techniques and  ML-based approaches is presented in Tables \ref{CNN_Forensics.tab} and \ref{tab.adv.ImgFor}, respectively.
		\begin{table*}[h!]
		\renewcommand\arraystretch{2.0}
		\centering
		\begin{center}
			\caption{Summary of different CNN-based image forensics techniques}
			\resizebox{\columnwidth}{!}{%
				\begin{tabular}{| c | c | c |}
					\hline
					\bf{Reference} & \bf{Pros} & \bf{Cons}  \\
					\hline
					\cite{ChenMFwithCNN} & \makecell{Median filtering detection, particularly\\ in cut and paste manipulation.} & \makecell{Small image size detection with \\ low quality factor.} \\
					\hline
					\cite{StammCNNuniv} & \makecell{Different manipulation detection such as \\blurring, median filtering, and resizing.} & \makecell{Problem with small window size, \\ for example median $3 \times 3$.} \\
					\hline
					\cite{BayarCNNUnivTIFS} & \makecell{Extended approach of \cite{StammCNNuniv}.} & \makecell{Mis-matched dataset.} \\
					\hline
					\cite{niu2019primary} & \makecell{Compute primary quantization matrix \\ in DJPEG.} & \makecell{Tampering detection.} \\
					\hline
									
				\end{tabular}
				\label{CNN_Forensics.tab}
			}
			
		\end{center}
		
	\end{table*}
\end{itemize}
\begin{table*}[h!]
	\renewcommand\arraystretch{3.0}
	\begin{center}
		\caption{Summary of existing ML-based image forensic approaches}
		\begin{tabular}{| c | c | c |}
			\hline
			\bf{Detection task}  & \bf{SVM} & \bf{CNN} \\
			\hline
			\bf{Double JPEG Compression (DJPEG)} & \cite{chen2008machine,Ehsan2015JPEG,Milani2012DiscJPEG} & \cite{niu2019primary} \\
			\hline
			\bf{Contrast Enhancement} &	\cite{Eusipco2017Ehsan,IWBF2018Ehsan}& \cite{ICIP2018Ehsan,Xiao2019Contrast}	\\
			\hline	
			\bf{Splicing and Copy-Move} &	\cite{Karmakar2018,Amerini2011Copy}& \cite{barni2019copy,Abdalla2019CopyMoveFD}	\\
			\hline
			\bf{Photo Response Non Uniformity (PRNU)} &	\cite{Filler2008PatternNoise}& \cite{Davide2018Camera}	\\
			\hline
			\bf{Median Filtering} &	\cite{Rhee2017MedianFD}& \cite{ChenMFwithCNN,StammCNNuniv}	\\
			\hline
			\bf{Multi Purposes} &---& \cite{BayarCNNUnivTIFS,Divya2020CNNmulti}	\\
			\hline
		\end{tabular}
		\label{tab.adv.ImgFor}
	\end{center}
	
\end{table*}


\section{Review of image forensics literature} \label{section3}
Here, we will focus on addressing the following research questions:

	\begin{itemize}
	\item What is the trend of ML applications in image forensics? 
	\item How anti-CF techniques can be used to improve the security of ML against adversarial attacks? 
	\item What methods are available to improve the security of ML engines used in image forensics? 
	\end{itemize}
	
We searched the literature using the following keywords: \textit{(image forensics) OR (image-forensics) (counter forensics) OR (counter-forensics) OR (counterforensics) OR (anti-counter-forensics) OR (anti-counter forensics) OR (anti counter forensics)} on Google Scholar, IEEE Xplore Digital Library, Springer, and ScienceDirect. The searches were limited to the paper title, keywords, and for conference, journal, and magazine articles. 

We only included English-language studies that report on empirical and theoretical findings in image forensics using machine learning. We also considered patents. Table \ref{tab.criteria} describes our inclusion criteria. 
\begin{table*}[h!]
	\renewcommand\arraystretch{3.0}
	\begin{center}
		\caption{Inclusion criteria}
		\resizebox{\columnwidth}{!}{%
			\begin{tabular}{| c | c |}
				\hline
				\bf{Content}  & \bf{Expectation}  \\
				\hline
				\makecell{Empirical and theoretical papers related to image forensics} & \makecell{Papers focusing on forge image detection }.\\
				\hline
				\makecell{Empirical papers related to image security} & \makecell{Papers focusing on CF and anti-CF }.\\
				\hline
				\makecell{Machine learning on image security} & \makecell{\\Papers focusing on different \\ attacks and defensive methodologies \\}.\\
				\hline          
			\end{tabular}
			\label{tab.criteria}
		}	
	\end{center}
	
\end{table*}





\subsection{Analytical Discussion}
\begin{figure}[h!]
	\centering
	\includegraphics[width=0.8\columnwidth]{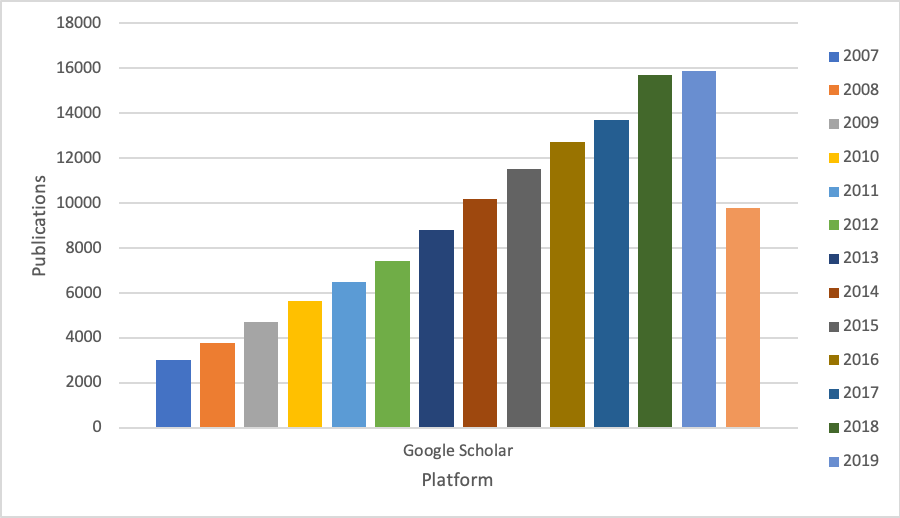}
	\caption{Number of image forensics articles}
	\label{Image_Forensics}
\end{figure}

\begin{figure}[h!]
	\centering
	\includegraphics[width=0.8\columnwidth]{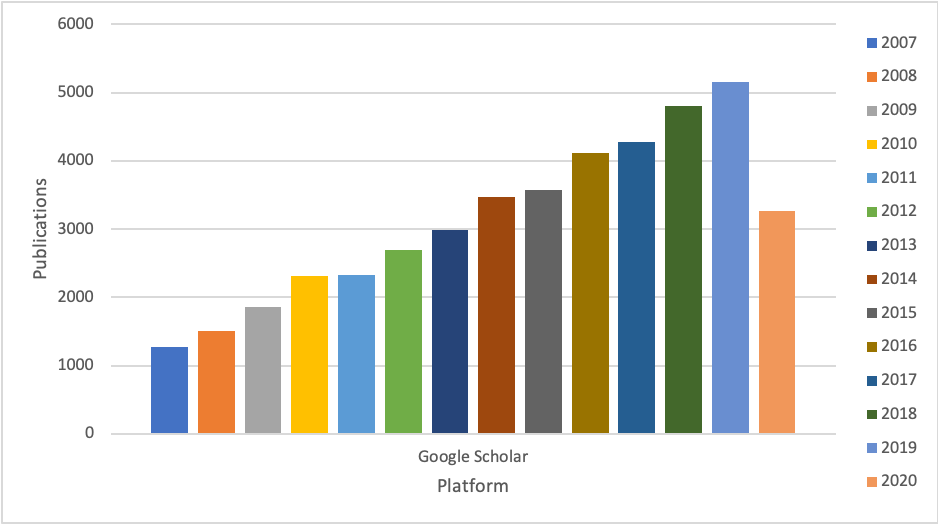}
	\caption{Number of CF articles}
	\label{CF1}
\end{figure}

\begin{figure}[h!]
	\centering
	\includegraphics[width=0.8\columnwidth]{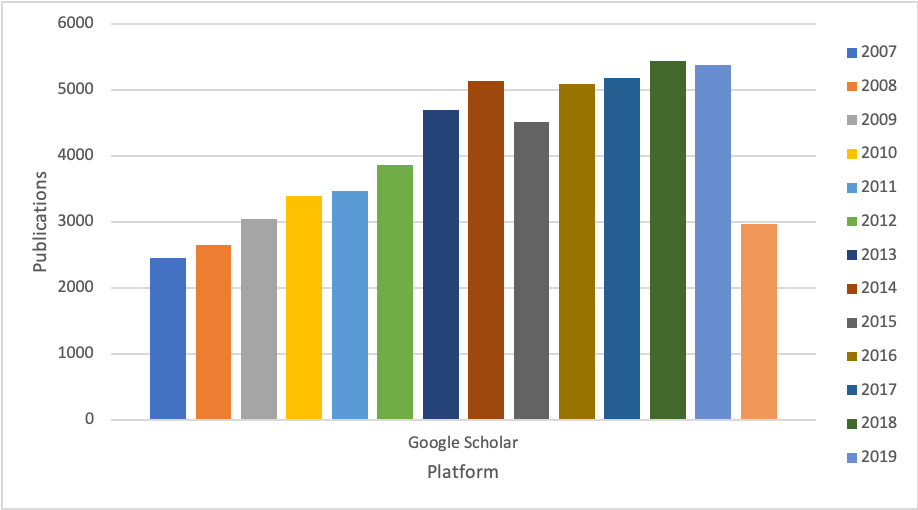}
	\caption{Number of anti-CF articles}
	\label{ACF1}
\end{figure}

Figure \ref{Image_Forensics} presents the number of image forensics articles included in our study, and Figures \ref{CF1} and \ref{ACF1} show the number of studies in image CF and image anti-CF respectively. 
%

\begin{figure}[h!]
	\centering
	\includegraphics[width=0.8\columnwidth]{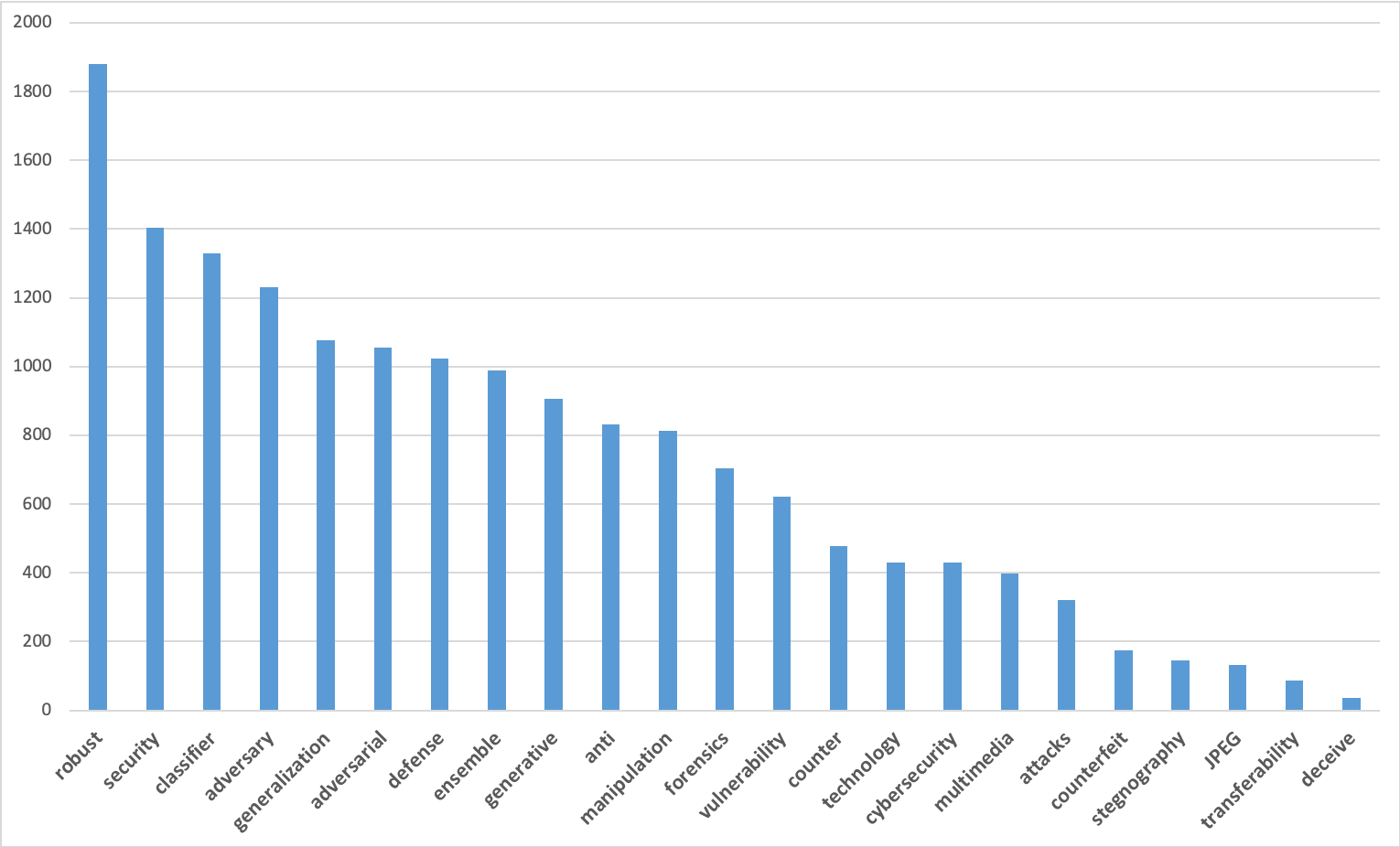}
	\caption{Keyword count in the included articles}
	\label{count}
\end{figure}

As shown in Figure \ref{count}, ``adversary'', ``robust'' and ``security'' are three popular keywords. This suggests a growing interest in studying the security of image files in the presence of an adversary.
%
%

\subsection{Knowledge Synthesization}
The primary keyword in Figure \ref{count} shows that there are important papers that are related to robustness, security, and classifier. 
	Based on this knowledge, the selected subjects only focused on the solutions to today's difficulties, and the few practical tools. 
	%
	Some practical security explications in ML, particularly for image forensics, only introduced a method for answering the specific problems in this area. Therefore, based on our knowledge of protecting the detectors against adversarial attacks requires a specific architecture that works against different attacks is infancy. Moreover, security and untrustworthy in ML techniques for forensic image detection become important. \\
	The studies showed that the security of ML techniques in the presence of an adversary plays an important role, and the development of solutions capable of improving security becomes necessary. In the following, to make this survey more focused, we considered three research questions.

\subsubsection{RQ1: what are the latest machine learning applications in image forensics? }


Image forensics' task has been deal with numerous algorithms based on statistical analysis and pattern recognition—lately, computing capabilities' improvement further attention in ML methods and particularly in deep learning, and have demonstrated their effectiveness in many image forensics competitions \cite{FERREIRA2020,Yang2020}. \\
Techniques based on ML/DL-based image forensics have been great attention in recent years. Methods based on DL have consistently achieved remarkable results on a series of tasks in image forensics such as Face GAN detection \cite{barni2020cnn}, Deepfake detection \cite{guarnera2020deepfake,verdoliva2020media}, Coding-based detection \cite{chen2008machine}, and so on.  Among them, CNN's are effective when dealing with various image tasks and utilized as a base for many image forensics techniques. One example related to median filtering detection is considering CNNs \cite{ChenMFwithCNN} and utilizing different detection tasks such as blurring, noise addition, resizing, rotation, etc.\\
Besides DL, in many complex tasks, researchers considered ML methods to address different forensics tasks, e.g., SVM. SVM-based in image forensics was recently utilized for detecting different forensics tasks, such as align and not-align DJPEG \cite{Ehsan2015JPEG}, global and local contrast enhancement \cite{IWBF2018Ehsan}, and so on. \\
Therefore, RQ1 plays a pivotal role in knowing new methodologies in adversarial image forensics, e.g., such as counter-forensics and anti-counter forensics. \\

\subsubsection{RQ2: how anti-CF techniques can improve the security of ML against adversarial attacks?}

Anti-forensics or counter-forensics are methods to make forensics algorithms fail by altering the image content or changing some forensic algorithm.  The main aim of the forensic community is to develop more robust forensic schemes. Recent studies in DL led to the development of generative adversarial networks (GANs) \cite{goodfellow2014explaining} that confirmed to be more effective for misleading many current image forensics approaches, referred to as adversarial examples. Adversarial examples employ small perturbations on the image to indue a system to make a wrong decision. The perturbation can be obtained by estimating the input image's loss function, such as the FGSM method \cite{goodfellow2014explaining}, Deepfool \cite{Moosavi2016DeepFool}, JSMA \cite{Paper16}, and so forth.  Guera et al., \cite{BestaAdv17} considered FGSM and JSMA adversarial attacks to mislead CNN camera model identification. Tondi \cite{TondiAttack} also considered a gradient attack on the pixel domain to create adversarial examples in the integer domain against CNN-based detection. According to the study, Most of the DL methods are fragile and vulnerable against adversarial attacks. The researchers report that even if the robustness of DL based method can be improved by retraining the classifier with adversarial examples, the resulting networks are still vulnerable against most powerful attacks.\\
Anti-counter forensics or counter-counter forensics have been extended as a defense methodology versus counter-forensics methods by developing image forensics' security in counter-forensics attacks.  While with the advent of adversarial attacks, various techniques have newly been developed in image forensics to defend versus adversarial attacks. Barni et al., \cite{Eusipco2017Ehsan} proposed a secure method to react against different post-processing by retraining unaware classifiers with the most powerful attacks (MPAs). Therefore, by understanding most CF attacks, the analyst can be developed a secure system to react against the most powerful CFs.

\subsubsection{RQ3: what methods are available to improve the security of ML engines used in image forensics?}

One of the hot topics in image forensics is developing methods for improving ML against different adversarial attacks. Most of the methods are tailored versus particular CF attacks. In this survey, we study the solutions suggested so far to counter CF attacks. Moreover, we distinguish the methods based on adversary-aware system and generally more secure detectors.  \\
In the first category, the analyst thought to be aware of the CF method the system is subject to and tries to develop a new algorithm to reveal the attack by looking at specific traces left by the tool. For instance, Barni et al. \cite{Eusipco2017Ehsan} proposed a detector based on a support-vector-machine (SVM) that fed with a large number of features to identify the traces left by D-JPEG in the presence of attacks.  With the advent of DL architectures, adversary-aware training has been widely used to improve DL models' robustness to adversarial examples \cite{goodfellow2014explaining}. Although these approaches effectively improve the security of ML in image forensics, it still works only against specific attacks.\\
In the second category, generally more secure detectors, most anti-CF methods are developed without considering developing a technique that analysts react appropriately against various kinds of attacks. Barni et al. \cite{Barni2020} proposed a detector against PK attacks through multiple-classifier architecture. The architecture is also known as a 1.5C classifier, consisting of one 2C classifier, two 1C classifiers, and a final 1C classifier. They access the performance of the 1.5C against three different manipulation tasks resizing, median filtering, and contrast enhancement.\\
In this survey, we study different methodologies proposed so far based on the above categories, and we address the pros and cons of methodologies.  
	%
	%

\section{Adversarial Image Forensics}
\label{section4}

With the advancement of forensics methods for retrieving information and tampering detection, in recent years, CF methods have been developed to prevent a correct detection. CF tools are usually effective due to the defects of forensic tools (most CF attacks take advantage of the weakness of the traces), which most of the time are not though to work under adversarial conditions. In real-world situations, the behavior of an adversary intending to stop the analysis cannot be ignored. Therefore, forensics analysis needs to up its game by improving its detection methods capable of working in an adversarial condition.

The task of improving forensics methods in an adversarial setting is even more challenging when ML-based, and in particular DL-based, forensic tools are adopted for the analysis. This challenge can be attributed to the inherent vulnerability of these tools, and the performance decreased on the different conditions during the testing phase, concerning those used for the training.
This calls for the improvement of more reliable ML-based forensic tools, which is called a new class of tools that can efficiently counter-CF attacks (referred to as anti-CF) while keeping the advantages of modern ML methods \cite{barni2018adversarial}.  The security of ML systems in the adversarial conditions is a common problem in comparison with many other security-sensitive applications. Hence, the use of comparable solutions should be considered to secure image forensic methods.  Some useful symbols used throughout this section are listed in Table \ref{symbol.tab}.
\begin{table}[h!]
	\renewcommand\arraystretch{2.5}
	\begin{center}
		\caption{ List of symbols}
		\begin{tabular}{| c | c | c | c |}
			\hline
			\bf{Symbol}  & \bf{Definition} & \bf{Symbol} & \bf{Definition} \\
			\hline
			$\phi$ & Forensic algorithm & $\hat{\phi}$ & Surrogate detector\\
			\hline
			$\phi_A$ & Refined detector & $l_i$ & algorithm parameters\\
			\hline
			$X$ & Feature Space & $D$ & Training data\\
			\hline
			$l_1$, $l_2$ & Unknown parameters & $\hat{l_1}$, $\hat{l}_2$ & Attacker guesses\\
			\hline
			$K$ & Real camera fingerprint & $\hat{K}$ & Estimation of $K$\\
			\hline	
			$X_{adv}$ & Adversarial image & $y$ & Class label\\
			\hline
			$c$ & Line search constant & $\varepsilon$ & Strength of attack\\
			\hline
			$R$ & Number of rows & $C$ & Number of columns\\
			\hline
			$\mathcal{D}_A$ & set of attacked images & $A^{*}$ & Optimum attack\\
			\hline
			$J$ & Loss & $\nabla $ &  \makecell{Gradient of the \\ cost function}\\
			\hline
		\end{tabular}
		\label{symbol.tab}
		
	\end{center}
	
\end{table}
%
	\subsection{Counter-Forensics and Anti-Counter-Forensics}
\label{CF-AntiCF}
\begin{figure}[h!]
	\centering
	\includegraphics[width=0.6\columnwidth]{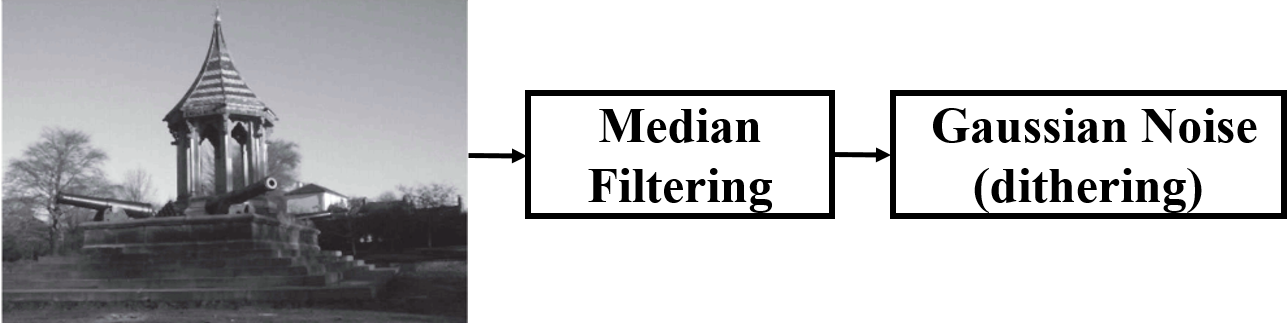}
	\caption{Scheme of the CF attack \cite{stamm2010undetectable}}
	\label{CF.Setup}
\end{figure}

CF or Anti-Forensics, refers to all the solutions proposed so far to bypass the forensic analysis. Early proposed models of CF methods were fairly simple, including applications of basic processing operators \cite{Bohme2008HidingTracesResample,cao2010anti,Stamm2011AntiForensicsCompress}. 
By adding noise dithering, the gaps in the histograms of DCT coefficients are decreased,  and the blocking artifacts are concealed by employing a smoothing operation, to hide traces of JPEG compression \cite{Stamm2011AntiForensicsCompress}. 
For instance, the most harmful attacks in image forensics related to the geometrical and CF attacks to eliminate the traces from previously JPEG images \cite{stamm2010undetectable} (see Figure \ref{CF.Setup}). In this case, the detector, in an unaware case, totally fails to detect. 

To conceal traces of contrast enhancement operation, the authors in \cite{cao2010anti} proposed a method that removes picks and gaps in the pixel histograms through dithering. To hide artifacts of resampling, image high frequencies are perturbed with noise while being resampled \cite{Bohme2008HidingTracesResample}. Such techniques can be easily circumvented by dedicated methods. 

The effective CF techniques can be devised by an attacker whenever he knows some information concerning the forensic algorithm. 
Such methods are assigned to as \textit{targeted attacks} and tailored to a particular algorithm.
CF methods are not perfect and leave traces on their own, that can be exploited by an informed analyst. The techniques developed by the analyst to defend against CF attacks are referred to as anti-CF techniques.

\subsection{Counter-Forensic Attacks}
\begin{figure}[h!]
	\centering
	\includegraphics[width=0.5\columnwidth]{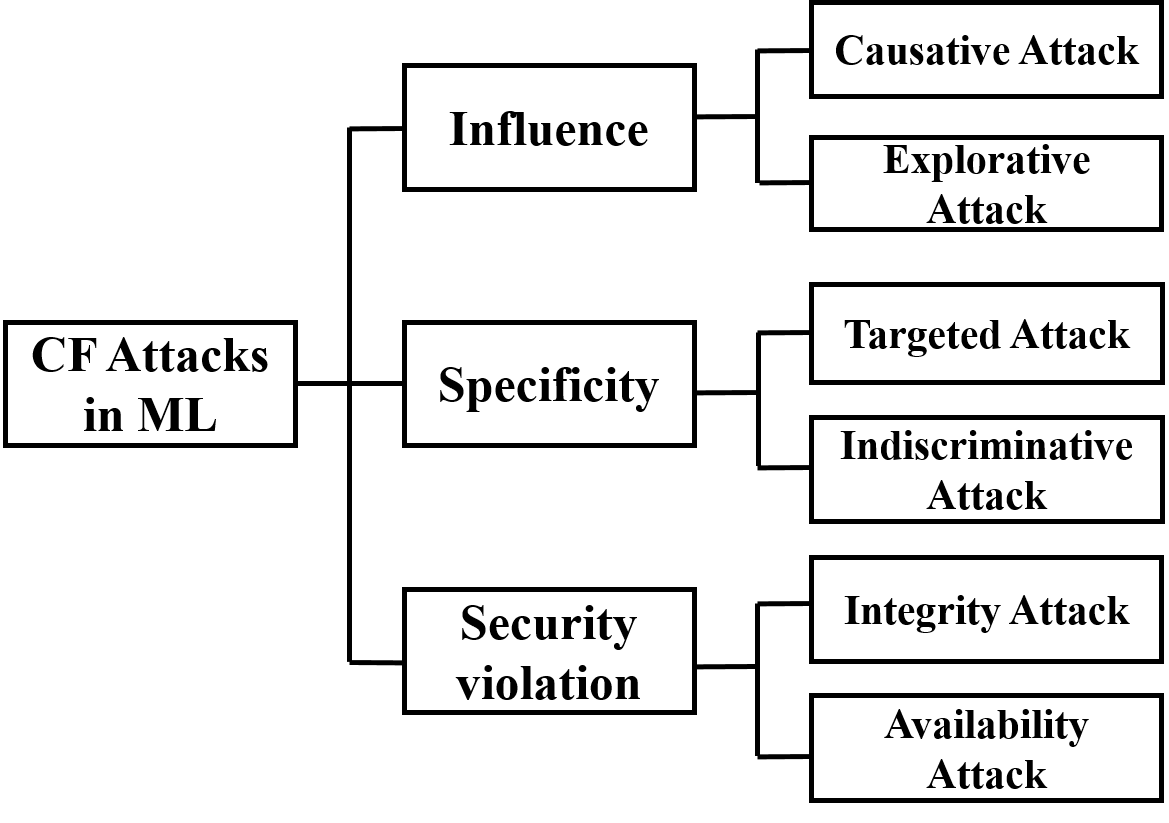}
	\caption{Classify the attacks in ML}
	\label{Figure1.Setup}
\end{figure}

We address useful terminologies to categorize the attacks in ML-based environments on influence, specificity, and security violations, as depicted in Figure \ref{Figure1.Setup}.
\begin{itemize}
	\item {\bf Influence:}
	\begin{itemize}
		\item {\em Causative}:  The attacker feeds malicious input to the training data. In this case, by feeding adversary samples during training, the learning parameters of models will be changed. Therefore, it can influence the performance of the classifiers. 
		\item  {\em Explorative}: The attacker feeds malicious input to the test data and never attempts to modify trained classifiers. The attacker tries to make a misclassification error through adversarial examples or seeks to obtain information from models.
	\end{itemize}
	\item {\bf Specificity:}
	\begin{itemize}
		\item {\em Targeted}: When the attacker attempts to decrease the performance of the classifier by the deception of a specific algorithm or specific group of samples.
		\item  {\em Indiscriminate}: When the attack has a more flexible goal, or when the attack is targeted to a class of algorithms rather than a specific algorithm.
	\end{itemize}
	\item {\bf Security violation:}
	\begin{itemize}
		\item {\em Integrity}: When the attack aims at letting malevolent samples be classified as normal, therefore attacker tries to increase false negatives error when classifying adversarial examples on the classifiers.  
		\item  {\em Availability}: When the attack aims at causing a classification error of any type, i.e., both a false negative and a false positive error, thus causing a Denial of Service (DoS) concerning pristine samples.
	\end{itemize}
\end{itemize}
Different kinds of attacks to the general ML system can be employed based on the above taxonomy, as done in \cite{Barni2020}. 

In the following, we study the general adversarial model utilized for CF attacks. Then, we address some examples of several attack models based on perfect and limited knowledge attacks. Moreover, we review different adversarial attacks against DL forensic techniques. \\

	\subsubsection{Counter-forensic attack model}
By following the terminologies in \cite{Barni2020}, an adversarial model is represented by specifying the assumptions about an attacker's goal, knowledge, and capability to corrupt the data or system. \\
\begin{itemize}
	
	\item {\bf Attacker's goal}\\ In this category, the attacker defines the kind of \textit{security violation}. The attacker aims to apply integrity attacks that induce false-negative error of the classifiers or employ an availability attack to induce errors of classifiers such as false-negative and false-positive error. We can classify CF into \textit{violation attacks} or \textit{evasion attacks} whenever the attacker modifies the manipulated images so that they are misclassified by the detector, either assumed as pristine ones or become misclassification errors. In doing so, the attacker normally wants to introduce a small distortion into the image to cross the boundary of a decision to minimize visual distortion while maximizing the loss function.\\
	
	\item {\bf Attacker's knowledge}\\ The knowledge of the attacker can be classified based on Perfect Knowledge (PK) or Limited Knowledge (LK), by examining the attacker knows features, training data, classifier architecture, learning parameters, and decision functions \cite{Biggio2013evasionAttack}. In the PK scenario, an attacker has the full knowledge regarding the forensic algorithm. This is the most convenient case for the attacker. Conversely, in the LK scenario, an attacker has only a few bits of information concerning the forensic algorithm, e.g., may not be aware of the exact algorithm or some of the parameters of the algorithm $l_i \in \mathcal{L}$. In ML-based techniques, the attacker knows only somewhat of information about the training data $\mathcal{D}$. Based on the available knowledge, the forensic algorithm specifies the \textit{specificity of attack} based on targeted or indiscriminate. Recently, \textit{universal} approaches are developed in such a way to be useful versus a whole class of forensic classifiers \cite{barni2014universal}. Barni et. al, \cite{barni2014universal} proposed a CF method that performs multiple compression undetectable based on investigating the histogram DCT coefficients. This approach removes double and multiple compression artifacts, while the visual quality of the image keeps high.
	
	\item {\bf Attacker's capability}\\ 
	In this case, the attacker can be controlling training or testing data, which referred to as \textit{influence} of the attack. We can interpret the impact of the attacker's capability whenever applied based on \textit{causative} or \textit{explorative} attacks. In the \textit{exploratory} attack, an adversary may modify test data, but can not alter training data. On the other hand, in a \textit{causative} attack scenario, the training process can be modified by an attacker, which referred to \textit{poisoning} attacks. Moreover, most of the CF attacks in nature are \textit{exploratory} \cite{Biggio2013evasionAttack}.
	
	The above explanation of the attack lets the attacker employ the threat model. Thus helps the analyst to design proper methods that can work in an adversarial situation.\\
	
\end{itemize}

\subsubsection{Perfect knowledge attacks}

\begin{figure}[h!]
	\centering
	\includegraphics[width=0.3\columnwidth]{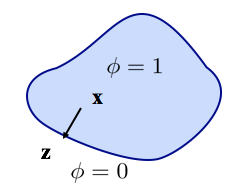}
	\caption{Scheme of Perfect Knowledge Attacks. The attacker knows $\phi$ = 1}
	\label{Figure2.Setup}
\end{figure}

The attacker in the PK scenario can employ his/her attack by relying on a forensic algorithm, denoted as $\phi$, and then lunch the attack to the target \cite{Bohme2013} (see Figure \ref{Figure2.Setup}).  
To keep a visual quality of the attack image, the attacker must induce an \textit{integrity violation}, which causes to false-negative decision error.
Particularly, the attacker requires to solve an optimization problem by looking at the image which is closest to the image under the attack.
Generally, an attacker needs to look at the image, which is resembling more to the image under the attack by solving an optimization problem.
Pasquini et al., \cite{Cecilia2014} employed a CF method to counter DJPEG compression detection by relying on First Significant Digits (FSD). Particularly, Comesana et al., \cite{Gonzalez2014FSD} shown the optimal attack against Benford's law-based detectors.  
Fontani et al., \cite{Barni2012HidingTraces} considered a method for hiding the median filtering artifacts and, in another case, counter SIFT-based copy-move detection, which is close to CF attacks. 

The optimum attack can be a gradient-based attack applied based on gradient-descent solutions when the detector is more complicated. To provide an example, Chen et al., \cite{Chen2017GraAttack} proposed a gradient-based attack employed on the pixel domain to counter SVM manipulation detectors on residual features. Similarly in \cite{TondiAttack}, a gradient analysis is introduced to counter forgery detection. When small perturbations applied to an image, the existing techniques tend to be canceled by pixels rounding, thus causing an attack useless. Therefore, the attack in \cite{TondiAttack} generates adversarial images with small perturbations to cause the classifier for the wrong decision. 

\begin{figure}[h!]
	\centering
	\includegraphics[width=0.4\columnwidth]{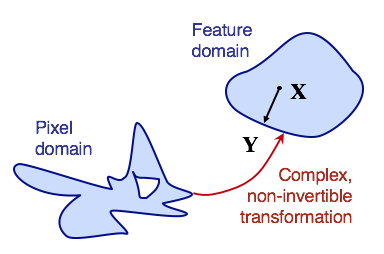}
	\caption{Complex non-invertable between pixel and feature domain}
	\label{Figure3.Setup}
\end{figure}

The main challenge with many PK scenarios is that most of the CF attacks are utilized in the feature domain. In this case, controlling the distortion will become a challenging task since the relationship among the pixel and feature domain is usually non-invertible \cite{Cecilia2014} (see Figure \ref{Figure3.Setup}).
Generally speaking, a suboptimum strategy is performed in two steps: First of all, it minimizes the distortion in the feature domain. Secondly, a new minimization is implemented in the pixel domain until it reaches a close desire attack \cite{Barni2012HidingTraces}.\\

\subsubsection{Limited knowledge attacks}
We consider the taxonomies were introduced in \cite{barni2018adversarial}, already employed to classify CF attacks based on \textit{universal attacks},  \textit{attacks on a surrogate detector}, and \textit{laundering-type attacks}.\\

\begin{itemize}
	
	\item {\bf Attacks on surrogate detector}\\ 
	In this case, the knowledge of the attacker for the algorithm $\phi$ is limited, and maybe he/she is aware of the $X$, the feature space, but the attacker does not know all the parameters of the algorithm $\mathcal{L}$ and also training set $\mathcal{D}$ in the case of ML-based techniques. The concept is shown in Figure \ref{Surrogate.Setup}.
	\begin{figure}[h!]
		\centering
		\includegraphics[width=0.5\columnwidth]{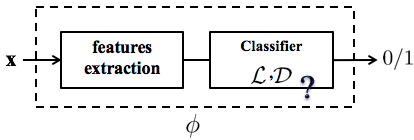}
		\caption{Scheme of the attack on the surrogate model}
		\label{Surrogate.Setup}
	\end{figure}
	In this case, the attacker creates a \textit{surrogate} detector ($\hat{\phi}$) and tries to guess the parameters he/she does not know based on whatever information that is available to him/her. Afterward, the attacker applies the CF by performing the attack versus ($\hat{\phi}$), expecting the attack will work against the real detector, which is commonly referred to as attack transferability \cite{PaperTransf16,Ehsan2019Transf}.  Recently, researchers prove that most of the MLs are fragile and vulnerable against adversarial attacks. Papernot et al., \cite{PaperTransf16} confirm that the attack on the source network entirely transferable to target networks. Whereas, Barni et al., \cite{Ehsan2019Transf} argues that adversarial attacks are not transferable from source to target in image forensics application. For instance, let $l_1$, and $l_2$ be the unknown parameters, then the attacker tries to guess the parameters $l_1$, and $l_2$ though his/her parameters are $\hat{l_1}$ and $\hat{l}_2$  and $\hat{\phi}$ where $\hat{\mathcal{L}} = \{\hat{l_1}, \hat{l}_2, l_3, l_4 ...\}$. The effectiveness of the method afterward is evaluated against the $\phi$ detector \cite{barni2018adversarial}.

An example of an attack on the surrogate model is fingerprint-copy attacks for PRNU camera identification. For the attacker, $K$ a real camera fingerprint is unknown, and then he/she tries to guess the parameters $K$ and $\hat{K}$ based on available images. Many attacks on ML-based detectors fall into this scenario.  
In fact, the attacker has limited access to the $D$, but he/she is aware of the architecture \cite{barni2018adversarial}. Then, the attacker builds $\hat{\mathcal{D}}$ as another dataset with the same distribution of $D$ and then employs it inside the pristine one. In other words, the attack duplicates information from the detector $\phi$ \cite{Chen2017GraAttack,Biggio2013evasionAttack}. Chen et al., \cite{Chen2017GraAttack} applied a pixel domain attack against ML forensic image detectors by considering a large dimensionality feature vector. They considered uncompressed images for this scenario because compressed images tend to erase the traces introduced in the pixel domain. Thus, attacks on compressed images need to go more inside the decision region, leading to high distortion on the image. More specifically, Biggio et al., \cite{Biggio2013evasionAttack} proposed a technique for evading classifiers via discriminant functions, which is effective in PK and LK scenarios.

	\item {\bf Universal attacks}\\
	In this category, the attacker does not know exact statistics performed by the analyst but is aware of the $X$ features; therefore, the attacker employs an effective attack against $\phi'$ inside the class of $\phi$ ($\phi$ equal to {$\phi(\mathcal{L}', \mathcal{X}), \forall \mathcal{L}'$}). Figure \ref{Universal.Setup} shows the universal attacks' procedure.
	\begin{figure}[h!]
		\centering
		\includegraphics[width=0.5\columnwidth]{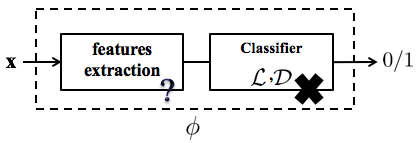}
		\caption{Scheme of the universal attacks}
		\label{Universal.Setup}
	\end{figure}
	Some of the examples are applied based on first-order statistics of the image versus the class of detectors to fool contrast enhancement \cite{Gonzalez2013CF} . 
For instance, Alfaro et al., \cite{Gonzalez2013CF}, considered a method based on the first-order statistics in the DCT domain to deceive DJPEG detection.\\

\item {\bf Laundering-type attacks}\\ 
The attacker only knows quite general and limited knowledge regarding the algorithm (see Figure \ref{Lundering.Setup}); then the attacker employs basic processing operation to erase the traces left by CF, such as considering filtering, resampling, and other processing. Moreover, the target of the attack does not limit only to a specific class or a detector. 
	\begin{figure}[h!]
		\centering
		\includegraphics[width=0.5\columnwidth]{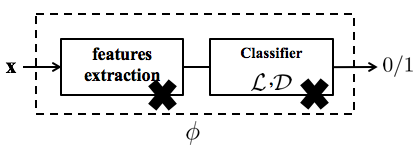}
		\caption{Scheme of the laundering attacks}
		\label{Lundering.Setup}
	\end{figure}

	Early laundering-type attacks have been developed against resampling detection \cite{Bohme2008HidingTracesResample}, S-JPEG compression \cite{Stamm2011AntiForensicsCompress}, and contrast adjustment \cite{cao2010anti}, to name a few as pointed out in Section \ref{CF-AntiCF}. In the literature, such attacks are usually related to targeted attacks, then they could be included in the PK scenario. The reason why these approaches such as contrast adjustment \cite{cao2010anti}, double JPEG compression \cite{Stamm2011AntiForensicsCompress}, are not included in PK attacks by \cite{barni2018adversarial} is the knowledge of these algorithms which is only marginally employed with basic processing which turns out to be enough for the attack purpose without any optimization.
	In most cases, the algorithm $\phi$ only utilized to prove the attack's effectiveness and not to guide the attack. 
	%
	
	Generally speaking, the implementation of such CF attacks is much easier than most PK attacks because the image distortion can be easily controlled by the attacker. \\
	
\end{itemize}

\subsubsection{Attacks on deep learning-based image forensics}

CNN architectures are capable of learning complex forensic features directly from input data or directly from an image. In contrast, an intelligent attacker can utilize this property and employ the most powerful CF attacks. 
In recent years, researchers in the ML area have found that the DL methods could be vulnerable and fragile against adversarial attacks, including  the forensics area. 
An attacker can create modified images and force misclassification errors due to the difference in the space of the inputs and images used to train the CNN, referred to as \textit{adversarial examples} \cite{szegedy2013intriguing}. 

Adversarial examples utilize the above property by applying small perturbations on the image to induce a system to make an incorrect decision. 
In this case, an adversarial image is generated, which is visually indistinguishable from the pristine one and misclassified by CNN. 
Adversarial perturbation can be obtained by estimating the gradient of the loss function of the input image, such as the Fast-Gradient-Sign-Method \cite{goodfellow2014explaining}, and DeepFool \cite{Moosavi2016DeepFool}.
Also, other attacks can be achieved by using iterative methods, including the Jacobian-Based Saliency Map Attack \cite{Paper16}, or by box constrained L-BFGS \cite{szegedy2013intriguing}. Adversarial examples to DL models are applied at testing time, which belongs to the category of explorative attacks and PK scenario. Adversarial examples in the PK scenario is often referred to as the \textit{white-box} scenario in DL literature. Whereas, the LK scenario is related to as a \textit{gray-box} scenario \cite{Zheng2018DNNrobustDetect}. In a realistic and challenging scenario, \textit{black-box} attacks are also considered in DL literature when the internal details (e.g., parameters) are not accessible to the attacker. For this reason, the attacker may employ several queries to gain internal details.

Recently, CF methods have been developed against CNNs in image forensics. Guera et al., \cite{BestaAdv17} considered the CF attack to deceive a camera model identification by considering FGSM and JSMA to mislead CNN-based camera model identification. 

Rounding process to integers is sometimes enough to wash out the perturbation and make an adversarial example ineffective. Tondi, \cite{TondiAttack}, applied a gradient attack on the pixel domain to create adversarial examples in the integer domain against a CNN-based detection.

The transferability of adversarial examples is the main concern in the security of ML-based image forensics, particularly in CNNs. Barni et al., \cite{Ehsan2019Transf} proved the adversarial examples are not transferable against CNN-based detectors. Whereas, Li et al., \cite{Li2020IncreasedconfidenceAE} proposed a new method  to improve the strength of the attack to assess the transferability of adversarial examples. As a result, they created a strong attack by enhancing the confidence of the misclassification.

In the following, we review recent adversarial attacks applied to ML, particularly on DL. The study mainly deals with methods that try to deceive the deep neural networks.\\
\begin{itemize}
	
	\item {\bf L-BFGS adversarial attack}
	\label{L-BFGSS}
	
	The authors in \cite{szegedy2013intriguing} proposed adversarial examples against the neural network in 2014, and they generated adversarial examples using box-constrained LBFGS. This method looks for an adversarial image ($X_{adv}$) with regard to input image $X$ under the $L_2$ distance, yet the label is different from the classifier. In $X$, we consider y as a ground truth class, then $X_{adv}$ is s such that $l(X_{adv}) \neq y$, that is, $l(X_{adv}) =  1 - y$, where $l()$ indicates the class label (then, $l(X_{adv}) = 0$ if $y = 1$). The constrained minimization problem is a challenging task to solve but can be formalized as follows:
	\begin{align}
	\min_{\hat{X}} & \quad ||X - \hat{X}||_2^2, 
	\quad l(\hat{X}) = 1 - y,
	\end{align}
	An approximately optimum solution can be used to solve the relaxed problem \cite{szegedy2013intriguing}: 
	\begin{align}
	\label{BFGS}
	\min_{X'} & \quad c ||X - \hat{X}||_2^2 -  J_{\theta}(X,y),
	\end{align}
	which $J_{\theta}(X,y)$  referred to as loss function such as neural network cross-entropy with parameter $\theta$ and scaler $c$.
	Moreover, the gradient-descent algorithm is utilized to \eqref{BFGS} to solve the problem. The gradient-descent algorithm is applied to solve (\ref{BFGS}). 
	To find the constant $c > 0$ line search is considered at a minimum distance to yield an adversarial example.
	Parameter $c$ is solved with different values through the bisection search of other techniques for one-dimensional optimization. 
	Based on the above formalization, an adversarial attack can compute perturbations and then apply to $X$ to fool a neural network. In this case, the human visual system can not distinguish between  $X_{adv}$ and a clean image of $X$. An example of this scenario is depicted in Figure \ref{LBFGS.Setup}.
	\begin{figure}[h!]
		\centering
		\includegraphics[width=0.5\columnwidth]{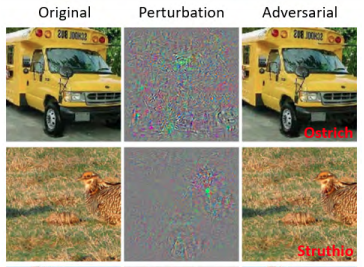}
		\caption{An example of adversarial examples is generated for AlexNet \cite{LeCun1989,szegedy2013intriguing}}
		\label{LBFGS.Setup}
	\end{figure}
	\\
	
	\item {\bf Fast Gradient Sign Method}\\
	Goodfellow, Shlens and Szegedy \cite{goodfellow2014explaining} proposed a fast, suboptimum adversarial attack, called FGSM. Given an image $X$, the FGSM method is formalized by
	\begin{align}
	\label{FGSM}
	\hat{X} =  X + \varepsilon * \text{sign}(\nabla_X J_{\theta}(X,y))
	\end{align}

	where a parameter $\varepsilon$ in \eqref{FGSM} referred to as the strength of the attack, which has a small parameter that leads to undetectable.   
	Intuitively speaking, this method determines in which direction the pixels should be modified by considering the gradient of the loss function. Figure \ref{FGSM_Fig.Setup} shows an adversarial example that is generated by a Fast Gradient Sign Method attack \cite{goodfellow2014explaining}.
	\begin{figure}[!htb]
		\centering
		\includegraphics[width=0.8\columnwidth]{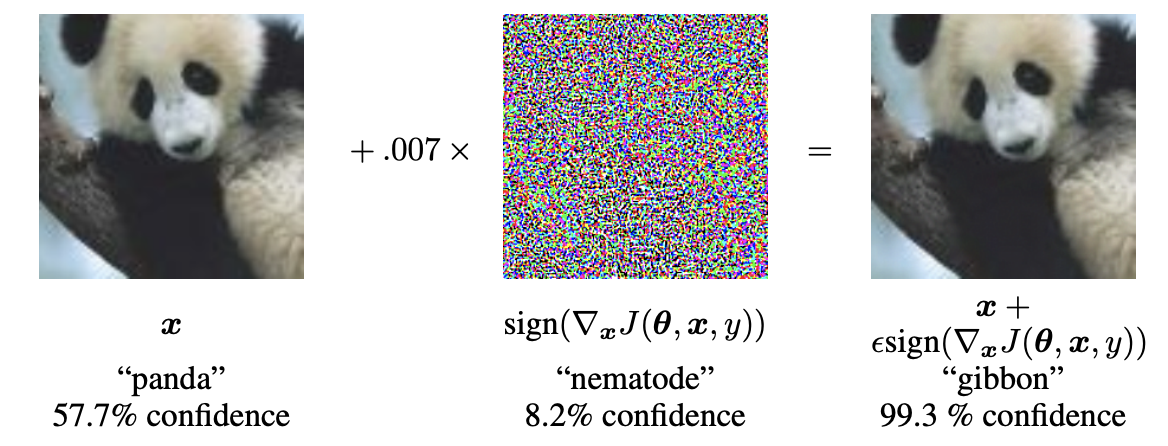}
		\caption{An example of FGSM applied to GoogLeNet on ImageNet \cite{goodfellow2014explaining}. Here the strength of the attack $\varepsilon$ considered as 0.007}
		\label{FGSM_Fig.Setup}
	\end{figure}
	In comparison with the previous attack in Section \ref{L-BFGSS}, L-BFGS, the FGSM adversarial attack is devised to be optimum and fast and has more efficient computation, yet, it does not create a close-to-minimal adversarial perturbation. Another difference with respect to L-BFGS, which is optimized for the $L_2$ distance metric, the FGSM method is optimized for the $L^{\infty}$  distance metric.
	
	Sometimes, an attack based on \eqref{FGSM} fails to get the adversarial images. So, another less suboptimum attack can be achieved by considering the \textit{iterative} version of the attack. In each iteration $i + 1$, the adversarial perturbation is obtained as specified in \eqref{FGSM}, and the image is updated as
	\begin{align}
	\label{I-FGSM}
	X_{i+1} = X_{i} + \varepsilon * \text{sign}(\nabla_X J_{\theta}(X_{i},y))
	\end{align}
	Moreover, in the direction of the gradient sign, the attack is employed iteratively $i$ with a small $\varepsilon$.\\
	
	\item {\bf Jacobian-based Saliency Map Attack}\\
	JSMA adversarial attack is a greedy iterative method relying on the forward propagation, proposed by \cite{Paper16}. This attack is optimized under the $L_0$ distance metric. 
	A saliency map is computed in each iteration for the pixels that have a high contribution to the classification, and then based on the saliency map, sensitive pixels are modified by the parameter of $\theta < 1$. Hence the modification of the pixels being,
	\begin{align}
	\label{JSMAA}
	\theta \cdot (\max(X_i) - \min(X_i))
	\end{align}
	The iterative process ends when either 1) adversarial examples are misclassified, 2) the attacker succeeds by getting an adversarial $X_{adv}$ such that l$l(X_{adv}) \neq y$, or 3)  whenever an adversarial image cannot be found for a given maximum $L_0$ distortion (sample of saliency map is illustrated in Figure \ref{SaliencyMap.Setup}).
	\begin{figure}[h!]
		\centering
		\includegraphics[width=0.3\columnwidth]{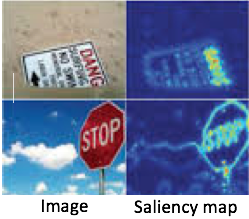}
		\caption{An example of Saliency Map}
		\label{SaliencyMap.Setup}
	\end{figure}
	\\
	
	\item {\bf Projected Gradient Descent}\\
	PGD adversarial attack determines the perturbation under constraints $L^{\infty}$ that maximizes the loss function \cite{madry2017towards}.
	
	In each iteration $i+1$, $X$ first is updated according to some rule $X_{i+1}$. Then, the image is projected onto the space having constrained $L^{\infty}$  distortion with the maximum distortion set to value $\alpha$, that is,
	\begin{align}
	\label{PGDD}
	X_{i+1} = \Pi_{X+\alpha}(X_{i+1})
	\end{align}
	In \eqref{PGDD} $\Pi$ is the projection operator. PGD is an iterative extension version of the FGSM and is similar to the I-FGSM. 
	
	Based on \cite{kurakin2016adversarial}, PGD adversarial attack is implemented as follows. Multiple times a gradient sign attack is employed with a small step size  $\varepsilon$ ($\varepsilon < \alpha$); 
	%
	\begin{equation}
	[X_{i+1}]_{R, C} = \text{clip}([X_{i+1}]_{R, C}),
	\end{equation}
	for pixel $(R,C)$, where $X_{i+1}$ uses the equation of \eqref{I-FGSM}.\\
	
	\item {\bf Attack on One-Pixel}\\
	Another scenario to deceive a classifier is based on the one-pixel attack.
	Su et al., \cite{DBLP:journals/corr/abs-1710-08864} demonstrated can successful deceive three different models by modifying one pixel per image through the concept of Differential Evolution \cite{Das2011Differential}. In their work, they received 70.97\% accuracy of the tested images with the average confidence of the network of 97.47\% on the wrong labels. Figure \ref{OnePixel.Setup} shows an example of the adversarial attacks on one-pixel.
	\begin{figure}[!h]
		\centering
		\includegraphics[width=0.4\columnwidth]{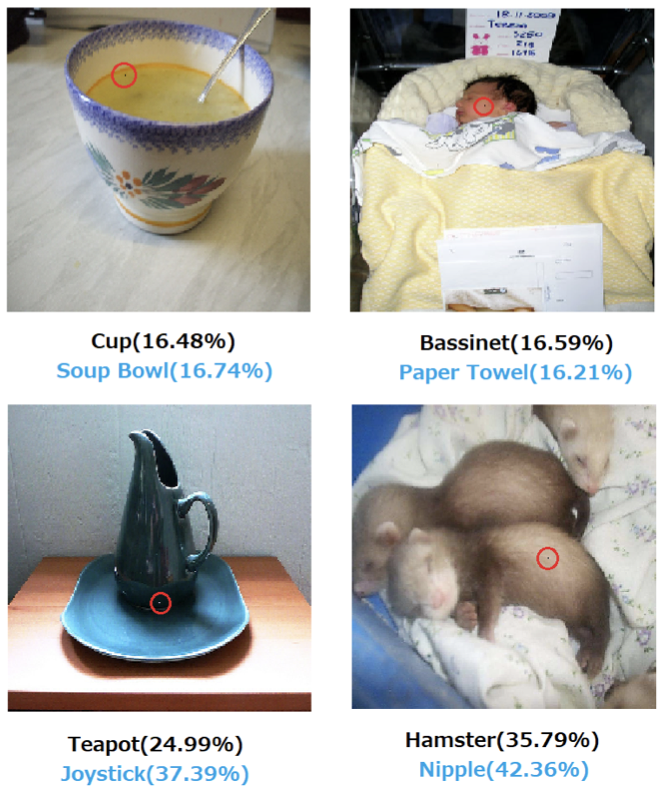}
		\caption{An example of one-pixel adversarial attacks \cite{DBLP:journals/corr/abs-1710-08864}}
		\label{OnePixel.Setup}
	\end{figure}
	For an image $X$, first, the vectors $R^5$ are generated, including RGB values for an optional candidate pixel and \textit{xy}-coordinates, then details are modified randomly to generate children vectors and including a child encounters and parent iteratively. 
	
	In comparison with previous attacks, differential evolution creates adversarial examples without any access to gradients, network parameters, and values. \\
	
	\item {\bf Carlini and Wagner Attacks} \\
	Carlini et al., \cite{Carlini16} explained that defensive distillation does not improve the robustness. To this end, they proposed three new attack algorithms and demonstrated that the defensive distillation completely fails against these attacks. They argue that their attacks are more effective concerning the three distance matrices $L_0$ , $L_2$ , and $L_{\infty}$  norms, as they called it the C\&W attack. Moreover, they proposed high-confidence adversarial samples in a transferability scenario that can defeat defensive distillation. The C\&W attacks can compute perturbations for a black-box scenario. Figure \ref{CW.Setup} shows examples of adversarial examples produced by C\&W adversarial attacks. \\
	\begin{figure}[!h]
		\centering
		\includegraphics[width=0.5\columnwidth]{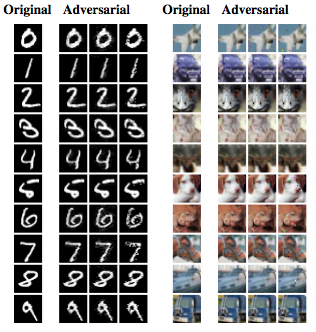}
		\caption{Example of C\&W adversarial attacks \cite{Carlini16}}
		\label{CW.Setup}
	\end{figure}

	\item {\bf DeepFool}\\
	Moosavi-Dezfooli et al., \cite{Moosavi2016DeepFool} initializes the image X that is restricted by the decision boundaries of a classifier. The region of the decision boundaries determines the class label of the image. In each iteration, a small vector applied to an image that is estimated by the boundary of the polyhedron. Then, in each iteration, the perturbations are applied to an image to measure the final perturbation based on the original decision classifier boundaries. The underlying algorithm is optimized for the $L_2$ and $L_{\infty}$ norms. \\
	
\end{itemize}

A summary of the main adversarial attacks recently applied to DL-based image forensics is provided in Table \ref{tab.adv.attacks}. 
%
\begin{table*}[h!]
	\renewcommand\arraystretch{2.0}
	\begin{center}
		\caption{Summary of diverse adversarial attacks to DL-based image forensics}
		\resizebox{\columnwidth}{!}{%
		\begin{tabular}{| c | c | c | c | c |}
			\hline
			\bf{Attacking Techniques}  & \bf{Attack Scenario} & \bf{Objective} & \bf{Learning} & \bf{Strength} \\
			\hline
			L-BFGS \cite{szegedy2013intriguing} & White box & Targeted & One shot & Medium \\
			\hline
			I-FGSM \cite{goodfellow2014explaining} & White box & Targeted & Iterative & Medium \\
			\hline
			JSMA \cite{Paper16} & White box & Non targeted & Iterative & Medium \\
			\hline
			One-pixel \cite{DBLP:journals/corr/abs-1710-08864} & Black box & Non targeted & Iterative & Low \\
			\hline
			C\&W \cite{Carlini16} & White box & Targeted & Iterative & High \\
			\hline
			DeepFool \cite{Moosavi2016DeepFool} & White box & Non targeted & Iterative & Medium \\
			\hline
			
		\end{tabular}
		\label{tab.adv.attacks}
}	
	\end{center}
	
\end{table*}
\begin{table*}[h!]
	\renewcommand\arraystretch{2.5}
	\begin{center}
		\caption{Summary of different counter-forensic techniques} 
		\resizebox{\columnwidth}{!}{%
			\begin{tabular}{| c | c | c | c |}
				\hline
				\bf{Knowledge}  & \bf{Reference} & \bf{Pros} & \bf{Cons} \\
				\hline
				Perfect & \cite{Cecilia2014} & Modifying the FSD histogram & \makecell{1) limit on distortion. \\ 2) Histogram reconstruction phase.} \\
				\hline
				Perfect & \cite{Gonzalez2014FSD} & \makecell{Optimal attack against histogram \\ where detection region is non-convex.} & \makecell{Considered only the MSE.} \\
				\hline
				Perfect & \cite{Barni2012HidingTraces} & \makecell{Conceal traces of median filtering.} & \makecell{Degrade the performance against \\ JPEG compression.} \\
				\hline
				Perfect & \cite{Chen2017GraAttack} & \makecell{Attack against SVM of global image\\ manipulation based on gradient descent.} & \makecell{With JPEG compression attacker \\needs to apply more distortion.} \\
				\hline
				Limited & \cite{PaperTransf16} & \makecell{\textbf{Attacks on surrogate detector:} \\ adversarial sample transfer across MLs.} & \makecell{Transferability can be improved by \\increasing the attack strength.} \\
				\hline
				Limited & \cite{Ehsan2019Transf} & \makecell{\textbf{Attacks on surrogate detector:} \\ Apply the scenario \cite{PaperTransf16} in image forensics.} & \makecell{Integer domain.} \\
				\hline
				 Limited & \cite{Gonzalez2013CF} & \makecell{\textbf{Universal attacks:} \\ fool a histogram-based forensics detector.} & \makecell{Can not applicable to other histogram \\ based detectors.} \\
				 \hline
				 Limited & \cite{Bohme2008HidingTracesResample} & \makecell{\textbf{Laundering attacks:} \\ overcome resampling detection.} & \makecell{Proposed methods are not detectable \\ with other existing forensic methods.} \\
				 \hline
				 Limited & \cite{Stamm2011AntiForensicsCompress} & \makecell{\textbf{Laundering attacks:} \\ remove JPEG artifacts.} & \makecell{Designed for specific scenario.} \\
				 \hline

			\end{tabular}
			\label{tab.CFF}
		}	
	\end{center}
	
\end{table*}
%
\subsubsection{Generative Adversarial Networks (GAN)}

With adversarial examples, CF scholars in the image forensics community have started studying Generative Adversarial Networks (GANs) \cite{Goodfellow2014GAN}. GANs are developed to generate generative models by imitating the distribution of training data. This is employed iteratively to make a min-max game between two-players, that is to say, discriminator train iteratively to distinguish between real and generated examples to deceive the discriminator. Newly, GANs have been employed as a CF attacks. For instance, Kim et al., \cite{Kim2017CNNantiMedian} present median filtering CF attacks by considering GAN network, which can effectively erase the traces of median filter images. Bonettini et al. \cite{bonettini2020use} considered Benford's law to discriminate generated images from pristine ones. Benford's law shows the distribution of the most important digit for DCT coefficients.
\subsubsection{Summary of counter-forensics attacks}

We summarize the different CF methods applied to an image to bypass the forensic analysis. Particularly, the problem of CF is discussed, and the related prior art is shown. As shown in Table \ref{tab.CFF}, an attacker with a perfect knowledge can make the attack by relying on the knowledge of the forensic algorithm. One the other hand, in a limited knowledge scenario, the attacker knows only some parts about the forensic algorithm, which means that it does not know the specific algorithm or maybe parameters.


\subsection{Anti-Counter Forensics (Anti-CF)}

Anti-CF methods have been extended to react to CF attacks and restore the validity of the forensic analysis. Most of the anti-CF methods are tailored against specific CF methods. Anti-CF methods can be classified into two categories: \textit{adversary-aware detectors} which is the most common case in image forensics, and \textit{generally more secure detectors} \cite{barni2018adversarial} when the attacker designs a secure system which is intrinsically difficult to attack.\\

\subsubsection{Adversary-aware detectors}

This is the most common approach in Adversarial Image Forensics, particularly in ML approaches. 
The analyst assumed to be aware of the CF attack; therefore, design a new method to expose the attack by looking at specific CF footprints. Then the new algorithm $\phi_A$ is devised within conjunction with the original algorithm $\phi$.

\begin{figure}[h!]
	\centering
	\includegraphics[width=0.35\columnwidth]{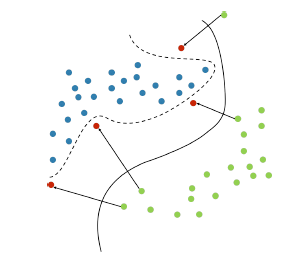}
	\caption{Refined the decision margin with attacked samples (red dots) and close more to pristine samples (blue dots) \cite{Eusipco2017Ehsan}.}
	\label{Adversary-Aware.Setup}
\end{figure}

Barni et al., \cite{Eusipco2017Ehsan} proposed a detector based on adversary-aware training to detect dangerousness CF attacks and other processing in the presence of DJPEG compression. 
They considered the \textit{Most Powerful Attacks} (MPAs), which degrade the classifier performance in an unaware case. Then, a detector retrained to identify images subject to MPAs should recognize milder processing and double compressed images. 
The example of this scenario is depicted in Figure \ref{Adversary-Aware.Setup}.
%
%
In other suited cases to ML-based is when the new algorithm is obtained by employing an adversary-aware version of $\phi_A$ and is utilized in place of $\phi$, indicating that the algorithm can be retrained by attack samples \cite{Eusipco2017Ehsan,ICIP2018Ehsan,IWBF2018Ehsan}.
In this way, the detector achieves the refined version of $\phi_A = \phi(\mathcal{L}, \mathcal{X}; \mathcal{D} \cup \mathcal{D}_A)$ by retraining the set of attacked examples $\mathcal{D}_A$. 
In general, this method works properly whenever the classifier can distinguish enough between pristine, manipulated, and adversarial examples. To provide an example, the authors in \cite{IWBF2018Ehsan} proposed a system capable of detecting contrast enhancement in the presence of JPEG compression. In this system, different SVM classifiers (unaware case), trained with the compressed images.  The second approach is to estimate QFs by utilizing the idempotency of JPEG images whenever QF cannot extract from the JPEG header.  Hence, different classifiers trained with different QFs separately, and based on the idempotency approach, the nearest QFs selected as a detector. Then, Barni et al., \cite{ICIP2018Ehsan} improved the approach in \cite{IWBF2018Ehsan} based on CNN detector for generic contrast adjustment against JPEG compression. In this system, robustness achieved by retraining unaware cases with JPEG examples with different QFs. Experimental results prove this system also can work under unseen tonal adjustments.

Furthermore, recently, forensic analysts extensively start to study how they can improve the robustness of DL based on adversary-aware training against adversarial attacks \cite{goodfellow2014explaining}. Want et al., \cite{Wang2020ICASSP} proposed a technique to prevent adversarial dangerousness attacks on DL based on multi-source and multi-cost schemes for improving defense performance, called Adversarially Trained Model Switching (AdvMS). Based on the AdvMS scheme, the first component multi-source alleviates the performance problem, and the second component multi-cost enhances the robustness.

The advent of Generative Adversarial Networks (GANs) develops many challenges in image forensics. GANs have been extensively employed to generate fake images. Identification of GANs-generated images is a significant forensics challenge, particularly for Image-to-Image translation and DeepFakes. Nataraj et al., \cite{Amit2019GANface} proposed a new method for identifying GAN images using a combination of different co-occurrence matrices. They consider matrices on three color channels RGB and train a model using CNN. The challenging task in this scenario is that the proposed methodology intrinsically is not robust against JPEG compression leading to the network being re-trained with different JPEG quality factors. In this work, they obtained 93.78\%, 91.61\%, and 87.31\% for quality factors of 95, 85, and 75, respectively. 


\subsubsection{Generally more secure detectors}

Most of the current anti-CF methods do not take into account the possibility of predicting the movement of the analyst and determining the probable course of action for each attack. When the attacker predicts that CF footprints are left by themselves, they devise more robust CF methods that leave less evidence, leading to a loop between CF and anti-CF techniques. 

A possible solution concerning this problem is to devise a system that resists against PK attack and inherently difficult to attack. In compered with \textit{adversary-aware training}, the analyst makes a system to counter several CF attacks, for instance, one possible approach is to consider higher-order statistics.
In this case, $\mathcal{X}$  is the set of features from the original algorithm, and $\mathcal{\hat{X}}$ is a larger feature space that estimates by the algorithm, where $\mathcal {\hat{X}} \supset \mathcal{X}$.
This method was implemented by Rosa et al., \cite{Rosa2015ContrastCF} by applying second-order statistics for contrast enhancement detection. The security of the proposed system in \cite{Rosa2015ContrastCF} evaluated the universal CF scheme \cite{barni2014universal}. They achieved three interesting facts: 1) This system is slightly better than the first-order detector; 2) CF attack methods recently applied against first-order statistics detectors are not functional versus second-order statistics; 3) Evaluate traces that left by an attack; it is easy in second-order statistics. Furthermore, a similar approach was proposed for countering S-JPEG, DJPEG, and local tampering anti-forensics in \cite{Singh2019JPEGAnti,chen2008machine}. 
Considering second-order statistics reveal CF attacks, helping the analyst to perform a more correct analysis. Another strategy to design a generally more secure system based on fusing the outputs of different forensic algorithms by looking at traces \cite{Fontani2014Countering}. Researchers borrowed this scenario based on Dempster-Shafer and investigated how Counter-Anti-Forensics (CAF) tools can be embedded in this system for classifying CAF traces from image forensic traces.

General speaking approaches referring to the category of \textit{generally more secure detector} look for solutions in the presence of the worst-case, for a given class of attacks. An example of such a technique can be found in \cite{barni2016adversary} for detecting DJPEG compression by re-training on $\mathcal{D}\cup {\mathcal{D}}_{A^{*}}$, where $A^{*}$ is considered as an optimum attack. In the following, we present an overview of recent works developed for improving the security of detectors.
\\
\begin{itemize}
	\item \textbf{Data Randomization} \\
	Randomization strategy is another method in this category that can be employed to enhance the robustness for forensic detectors on DL for general models and standard-based forensics \cite{Roli2016Evasion,Ehsan2020RDFS}. Zhang et al., \cite{Roli2016Evasion} they optimized the feature sets, which intrinsically become secure against a PK attack by considering feature selection method on adversarial samples that increase the security versus attacks at test time. Therefore, feature randomization plays an important role in the security-related application when a small set of features are considered to overcome the complexity or even enhance the performance of classification to tackle adversarial attacks. Barni et al., \cite{Ehsan2020RDFS} considered random feature selection strategy that can improve the security and robustness of forensic detectors for the general model and standard ML-based to mitigate the adversarial dangerousness attacks. The experiments prove that feature randomization strategy reducing the transferability of attacks, and increase the security of detection even in the presence of mismatch architectures. Moreover, such techniques also have been confirmed to be effective against PK attacks \cite{Chen2019Randomization}. 
\end{itemize}	

\begin{itemize}
	\item \textbf{Defense Layer}\\
	In another strategy adding a new defense layer in a network helps to counter adversarial attacks in the back-box setting. In this approach, the defense layer parameters assist in gaining protection versus adversarial attacks (see Figure \ref{DefenseLayer}) \cite{Akshay2020}.	\\
	\begin{figure}[h!]
		\centering
		\includegraphics[width=0.6\columnwidth]{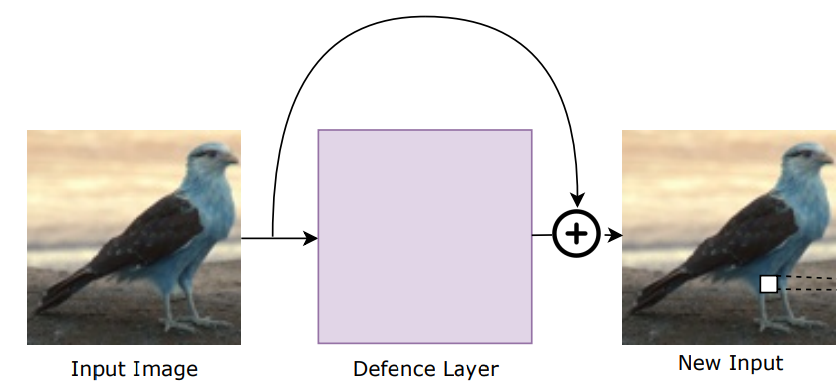}
		\caption{Defense layer architecture \cite{Akshay2020}}
		\label{DefenseLayer}
	\end{figure}
\end{itemize}

\begin{itemize}
	\item \textbf{One Class Classifier}\\
	One class classifier (1C) has been recently used by the forensic community to improve the security and robustness of image forensics, particularly in security applications.
	%
	%
	The main idea behind using autoencoders in image forensics is that they play as a role of 1C classifiers, which forgery data can be defined as an anomaly. 
	More in general, in many different applications, 1C modeling is famous for anomaly detection, when a statistical characterization under abnormal situations is not feasible. 
	%
	%
    1C combinations were devised for improving the security versus evasion attacks for adversarial anomaly detection. Moreover, 1C classification is an alternative approach for conventional multiclass algorithms that classify the examples based on several pre-defined categories, which referred to as \textit{open set} conditions. 
	In particular, in several forensic tasks and security-oriented tools, the \textit{open set} problem has been investigated so far in \cite{Wang2009SourceCamera}.
	Wang et al., \cite{Wang2009SourceCamera} considered SVMs multi-class and 1C to recognize different camera models. This technique is more robust against DJPEG images.
	In image security, 1C classifiers have also been developed in association with GAN to design detectors when malicious examples in training exist. Yarlagadda et al., \cite{Delp2018Satellite} considered this methodology for satellite images to evaluate their authenticity. In this architecture, GAN learns pristine satellite features, and 1C (SVM) is trained with those features. As illustrated in Figure \ref{Satellite}, the rational idea behind discriminator is to distinguish patches from a generator and pristine satellite images accurately. In contrast, the generator intends to deceive the discriminator by generating data near a pristine one.
	\begin{figure}[h!]
		\centering
		\includegraphics[width=0.8\columnwidth]{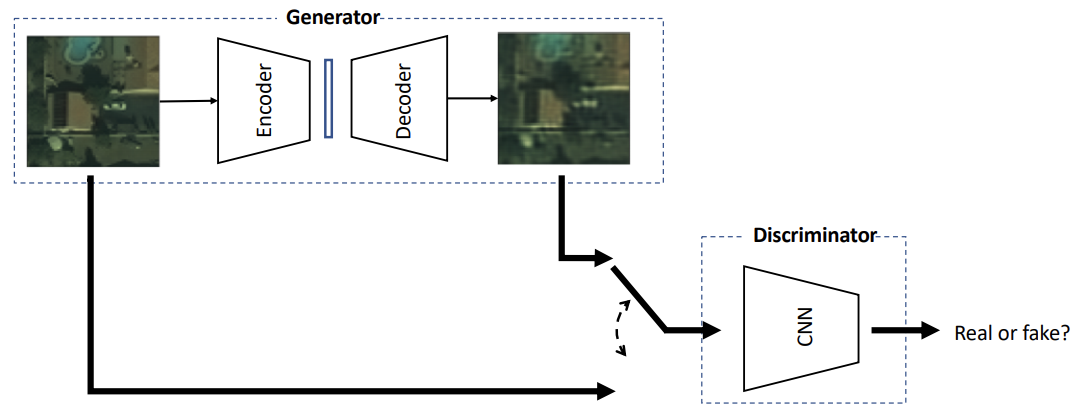}
		\caption{Architecture satellite Image foorgery detection \cite{Delp2018Satellite}}
		\label{Satellite}
	\end{figure}
\end{itemize}

\begin{itemize}
	\item \textbf{Ensemble Method} \\
Multiple architecture combination (1.5C) is another strategy to improve the security and robustness against PK attacks in image manipulation detection, which decreases damage caused by an attacker \cite{Barni2020}. 
The 1.5C architecture is achieved by a combination of 2C and 1Cs classifiers, runs simultaneously, and outputs of each classifier feed to the final 1C classifier.
The logical idea behind scenario 1.5C is as follows. 
2C classifiers always generalize entirely to the samples that were not represented correctly in training but achieved high accuracy and performance, especially in the absence of attacks.
Therefore, based on this problem, an intelligent attacker can use this property and applied the attack and fall into 'unseen' regions, which is illustrated in Figure \ref{Toy2C.Setup} \cite{barni2016adversary}. 
In Figure \ref{Toy2C.Setup}, the attacker aims to take manipulation samples (red triangles) and transfer to a pristine unseen region (blue dots). 
	\begin{figure}[h!]
		\centering
		\includegraphics[width=0.22\columnwidth]{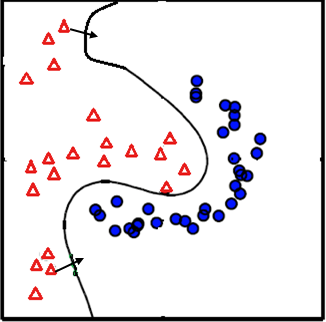}
		\caption{Scheme of 2C when the attacker transferred 'red triangles' to unseen 'blue dots' region \cite{Barni2020}}
		\label{Toy2C.Setup}
	\end{figure}
	By enclosing only one class, the system will become robust and secure against adversarial attacks.
	This behavior is depicted in Figure \ref{Toy1C.Setup}. 
	We see that moving samples from manipulated to the pristine region need a more distortion because of the adjacency of pristine samples to the decision region.
	\begin{figure}[h!]
		\centering
		\includegraphics[width=0.22\columnwidth]{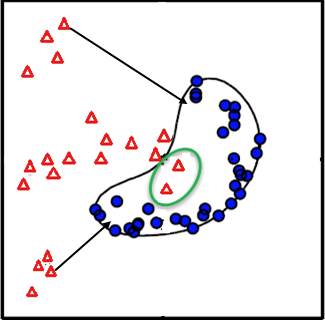}
		\caption{Scheme of 1C by defining a closed region \cite{Barni2020}}
		\label{Toy1C.Setup}
	\end{figure}
	To achieve a system that is intrinsically difficult to attack, the classifiers are combined, which has similar accuracy with respect to 2C and high accuracy with respect to 1C. 
	Based on Figure \ref{Toy15Cleg.Setup}, the acceptance region is well-shaped compared to 1C with similar accuracy with respect to 2C \cite{Barni2020}. \\
	\begin{figure}[h!]
		\centering
		\includegraphics[width=0.22\columnwidth]{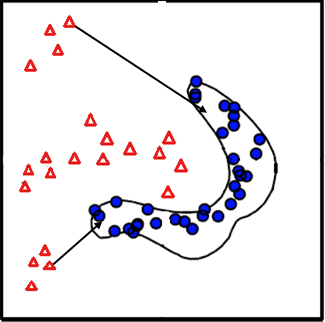}
		\caption{Scheme of 1.5C by defining a well-shaped decision region \cite{Barni2020}}
		\label{Toy15Cleg.Setup}
	\end{figure}
Recent advancements in technology have created realistic images via GAN as to be imperceptible from the pristine ones. Rana et. al. \cite{Rana2020DeepFake} proposed deep ensemble methods for detecting Deepfake videos by combining various DL classification models.
\end{itemize}

\begin{itemize}
	\item \textbf{GAN Defense} \\
	Lee et al., \cite{Lee2017GANdefensee} directly train a network besides the generator and tries to create the perturbation for that network. Thus, the classifier can correctly distinguish between pristine (clean) and manipulated one (perturb image). The proposed approach is robust against adversarial attacks such as FGSM \cite{goodfellow2014explaining}.  \\
	Defense-GAN is another methodology defending classifiers versus adversarial attacks by training the model with the distribution of unperturbed images \cite{Chellappa2018DefenseGAN}. The system tries to obtain a close output, which does not include the adversarial modifications. Therefore, this system can be employed as a defense technique against any attacks in the LK scenario. \\ While modern GAN can create fake images with high quality, re-construction of relations among color bands is expectedly difficult. Barni et al., \cite{barni2020cnn}  proposed a methodology for detecting GAN generated face images via co-occurrences analysis. 
\end{itemize}

	

Table \ref{Attack-Defense.tab} summarizes the various attacking and defensive ML-based techniques, and Table \ref{Attack-Defenseee.tab} presents experimental findings of the approaches described in Table \ref{Attack-Defense.tab}.
\begin{table*}[h!]
	\renewcommand\arraystretch{2.5}
	\begin{center}
		\caption{Summary of different anti-counter-forensic (anti-CF) techniques}
		\resizebox{\columnwidth}{!}{%
			\begin{tabular}{| c | c | c | c |}
				\hline
				\bf{Method}  & \bf{Reference} & \bf{Pros} & \bf{Cons} \\
				\hline
				\makecell{Adv-aware \\ detectors} & \cite{Eusipco2017Ehsan} & \makecell{Detect dangerousness attacks and \\ other processings in the presence of \\ DJPEG compression.} & \makecell{1) Degrade performance when $QF_1 > QF_2$. \\ 2) Localization detection.} \\
				\hline
				\makecell{Adv-aware \\ detectors} & \cite{Wang2020ICASSP} & \makecell{Prevent adversarial dangerousness attacks \\ on DL by employing multi-source and \\ multi-cost schemes.} &\makecell{Reduce the performance with large $\epsilon$.} \\
				\hline
				\makecell{Adv-aware \\ detectors} & \cite{Amit2019GANface} & \makecell{Identifying fake GAN images using \\ combination of different co-occurrence matrices.} &\makecell{Performance degrade on high quality images \\ such as StyleGAN1 or 2.} \\
				\hline
				\makecell{Generally more\\ secure detectors} & \cite{Rosa2015ContrastCF} & \makecell{Contrast enhancement detection by employing \\ second-order statistics.} &\makecell{Performance degrade against JPEG compression.} \\
				\hline
				\makecell{Generally more\\ secure detectors} & \cite{Singh2019JPEGAnti} & \makecell{Countering SJPEG, DJPEG, and \\ local tampering anti-forensics.} &\makecell{Localization detection.} \\
				\hline		
				\makecell{Generally more\\ secure detectors} & \cite{Fontani2014Countering} & \makecell{Fusing the outputs of different forensic \\ algorithms.} &\makecell{$QF1 < QF2 $.} \\
				\hline
				\makecell{Generally more\\ secure detectors} & \cite{barni2016adversary} & \makecell{Detecting DJPEG compression.} &\makecell{Works only with SVM.} \\
				\hline	
				\makecell{Generally more\\ secure detectors} & \cite{Ehsan2020RDFS} & \makecell{\textbf{Data randomization:}\\ improve the security in different \\forensic scenarios.} &\makecell{Adversarial examples can be transferable \\  by increasing strength of attacks.} \\
				\hline	
				\makecell{Generally more\\ secure detectors} & \cite{Akshay2020} & \makecell{\textbf{Defense layer:}\\ Counter adversarial attacks.} &\makecell{It cannot work in a white-box scenario.} \\
				\hline	
				\makecell{Generally more\\ secure detectors} & \cite{Barni2020} & \makecell{\textbf{Ensemble Method:}\\ Protect detectors against PK attacks.} &\makecell{Not to much robust against: \\1) Noise addition \\ 2) JPEG compression.} \\
				\hline	
				\makecell{Generally more\\ secure detectors} & \cite{Lee2017GANdefensee} & \makecell{\textbf{GAN Defence}:\\ Robust against adversarial examples using \\a GAN.} &\makecell{Problem with deep network.} \\
				\hline	

			\end{tabular}
			\label{tab.CFF_Anti}
		}	
	\end{center}
	
\end{table*}
%

\begin{table*}[htbp]
	\renewcommand\arraystretch{2.5}
	\centering
	\begin{center}
		\caption{Summary of various attacking and defensive techniques used in ML}
		\begin{adjustbox}{angle=90}
		\begin{tabular}{| c | c | c |c |c |c |}
			\hline
			\bf{\makecell{Attack-\\Defense}}  & \bf{Target} & \bf{Scenario} & \bf{Reference} & \bf{Charactristics} & \bf{Rational Idea} \\
			\hline
			Attack & Training & Poisoning& \makecell{\cite{Biggio2013poFace,Tianyu2017},\\ \cite{Biggio2014POi,Biggio2014MalwareClus}}& \makecell{
			Causative \\ Attack} & \makecell{Insert adversarial examples \\during the training phase.  \\Moreover, it can change the labels \\of the training set.}\\
			\hline
			Attack& Testing & Evasion & \cite{kurakin2016adversarial,Hu2017GANN,Biggio2013evasionAttack}& \makecell{Explorative \\Attack} &  \makecell{Attacks carried out at test time.}\\
			\hline
			Attack & Testing & Impersonate & \cite{Papernot2017PracticalBA,Carlini2016Hidden} & \makecell{Explorative\\ Attack} & \makecell{Apply adversarial examples \\ to simulate target ones or to \\confuse classifier decisions. } \\
			\hline
			Defense & \makecell{Training \\and Testing} & \makecell{Adversarial \\ Training} &  \cite{goodfellow2014explaining,Eusipco2017Ehsan,ICIP2018Ehsan,IWBF2018Ehsan, Wang2020ICASSP} & \makecell{Improve \\Robustness} &\makecell{Consider harmful attacks that \\degrade the performance of the \\classifier in the un-aware case \\and retraining the classifier \\with MPAs.} \\
			\hline
			Defense & \makecell{Training \\and Testing} & \makecell{Ensemble \\ Method} & \cite{Ehsan_Thesis} & \makecell{Improve  \\Robustness \\and Security }& \makecell{Combination of various classifiers to\\ alleviate the adversarial attacks. 
			} \\
			\hline
			Defense & \makecell{Training \\and Testing} & \makecell{Differential \\ Method} & \cite{Ehsan2020RDFS,Chen2019Randomization} & \makecell{Improve \\Security \\and Privacy} & \makecell{Apply random noise to data or \\employs randomized approaches.}\\
			\hline

		\end{tabular}
		\label{Attack-Defense.tab}
	\end{adjustbox}

	\end{center}
	
\end{table*}

\begin{table*}[htbp]
	\renewcommand\arraystretch{3.5}
	\centering
	\begin{center}
		\caption{Few experimental results of Table \ref{Attack-Defense.tab} in various techniques}
		\resizebox{\columnwidth}{!}{%
		\begin{tabular}{| c | c | c |c |}
			\hline
			\bf{Attack-Defense}  & \bf{Scenario} & \bf{Reference} & \bf{Results}  \\
			\hline
			Attack & Poisoning & \cite{Tianyu2017} & \makecell{Three different BadNets scenarios on traffic sign image \\(yellow square, an image of a bomb, and an image \\of a flower) misclassified more than 90\%.} \\
			\hline
			Attack & Evasion & \cite{kurakin2016adversarial} &  \makecell{With attack scenario FGSM with $\varepsilon$= 16, they achieved \\91.9\% accuracy for clean image and 67.7\% accuracy for an \\adversarial image. For an iterative version of FGSM with $\varepsilon$= 16, \\they obtained 89.6\% accuracy for clean image \\and 75.0\% for an adversarial image.}\\
			\hline
			Attack & Impersonate & \cite{Papernot2017PracticalBA} & \makecell{For DNN with id (A) at $\rho$ = 2 and $\rho$ = 6, they achieved \\30.50\% and 82.81\% accuracies respectively. In another \\example, for DNN with id (F), the at $\rho$ = 2 and $\rho$ = 6, they \\achieved 68.67\% and 79.19\% accuracies. } \\
			\hline
			Defense & \makecell{Adversarial \\ Training} &\cite{Eusipco2017Ehsan}& \makecell{ They applied different processing against refined classifier, \\which is re-trained with MPAs. Therefore, they achieved \\0.99 AUC for DJPEG, 0.98 for 1st order MPA, 0.91 for \\wavelet denoise, 0.98 for median filtering, 0.91 for clip limit \\adaptive histogram equalization (CLAHE), 0.92 for resizing 0.9 \\with bicubic interpolation, 0.92 for rotation 5 with bicubic\\ interpolation, 0.97 for zoom 1.2, 0.92 for crop align, \\0.93 for crop no align, 0.99 for a mirror, 0.98 for the blur, \\and finally 0.95 for seam-carving.}\\
			\hline
			Defense & \makecell{Ensemble \\ Method }& \cite{Barni2020} & \makecell{When the gradient attack employed on pixel domain \\against a two-class classifier (2C), they achieved zero \\percentage of misclassified of adversarial images for \\different detection tasks such as resizing, median \\filtering, and CLAHE.}\\
			\hline
			Defense & \makecell{Differential \\ Method} & \cite{Ehsan2020RDFS}& \makecell{Accuracy gain is about 76.30\% for random feature size $K$ = 30 \\when the FGSM adversarial attack applied on the network \cite{ICIP2018Ehsan}.} \\
			\hline

		\end{tabular}
		\label{Attack-Defenseee.tab}
	}
	
	\end{center}
	
\end{table*}
\subsubsection{Summary of Anti-Counter Forensics }

In this part, we summarize different anti-CF techniques into two categories \textit{adversary-aware detectors} and \textit{generally more secure detectors}. Particularly, we compare existing techniques in terms of pros and cons, as shown in Table \ref{tab.CFF_Anti}.  Table \ref{tab.CFF_Anti} shows that different techniques have their pros and cons in terms of the capability of detection, robustness, efficiency, and so on. Specifically, in \textit{adversary-aware detectors},  are good to react only against a specific case of attacks. On the other hand, in \textit{generally more secure detectors}, are more resistant versus CF attacks and then, more difficult to attack, particularly even in the PK scenario.

\section{Concluding remarks}
\label{sec.conc}
This article reviewed existing ML-based approaches for image manipulation detection in an adversarial setting, such as those based on binary manipulation classification. While a number of these approaches may be efficient in different forensic scenarios, they may be vulnerable or less effective against adversarial attacks. There have also been attempts to design image forensic tools for retrieving information and manipulation detection, as well as CF techniques to defeat ML-based image forensic approaches. 

We also observed that anti-CF approaches are generally tailored for specific CF attacks. For example, in the context of the Most Powerful Attacks (MPAs), MPA-aware detectors are designed to enhance the security of ML-based image manipulation against a specific class of attacks. In practice, when the MPA cannot be found analytically, a possible approach is to try to determine the MPAs experimentally, by looking at which attack degrades the performance \cite{Eusipco2017Ehsan}. 

In anti-CF, an aware-training technique is considered as a way to improve the robustness of the forensic tools against the JPEG laundering attack. We observed that image forensics tools may have poor resilience against JPEG compression. Compared to more complicated attacks, the JPEG laundering attack is easy to perform and does not require any information about the target detector. With reference to the contrast enhancement (CE) detection, the effectiveness of the JPEG-aware training has been evaluated for both ML (SVM-based) and DL (CNN-based) approaches in the literature \cite{Ehsan_Thesis,ICIP2018Ehsan,IWBF2018Ehsan}. There have also been attempts to design anti-CF systems that are generally difficult to attack by improving the security of ML-based image manipulation detectors \cite{Barni2020}, for example using randomization of the feature set \cite{Ehsan2020RDFS}. 

Recent investigations in DL have confirmed that adversarial examples present a certain degree of transferability, especially in the computer vision. Additionally, applying adversarial attacks can be efficient in Limited Knowledge (LK) scenarios. For example, Barni et al., \cite{Ehsan2019Transf} showed that adversarial examples against CNN-based detectors are not transferable between matched and mismatched architectures. Random feature selection (RFS) method is one possible approach to improve the robustness of forensic detectors to targeted attacks, particularly in DL.  In CNN, the RFS approach includes the selection of a random feature set derived from the network. Thus, the RFS strategy helps to alleviate the dangerousness of adversarial examples \cite{Ehsan2020RDFS}.

There remain a number of potential research directions, such as the following.
\begin{itemize}
	
	\item \textbf{\textit{Vulnerability and fragility of DL methods against adversarial attacks}} The reviewed literature shows that most DL techniques in image forensics can easily fool against various adversarial attacks. Although most of the current works focus on misleading DL on the task of classification, based on the studied literature, we can undoubtedly observe that DL techniques are fragile to adversarial attacks in general. Improving the robustness of the DL-based engines against adversarial learning techniques plays a pivotal role in image forensics. Newly, random selection approaches employes to increase the robustness of the forensic image detectors to targeted attacks. This approach extended based on DL features can help prevent attack transferability by modifying the detector's architecture or retraining the detector with random features. On the other hand, increasing the approach's effectiveness of this approach must be examined against a wide range of adversarial attacks.
	Moreover, identifying the robustness of the random deep feature selection scenarios when the adversary is aware of the defensive mechanism is another interesting avenue in future research. 
	
	\item \textbf{\textit{Adversarial samples generalize well}} The reviewed literature proves that one of the most common features in adversarial examples is that they transfer considerably between similar and different architectures. In contrast, the authors in \cite{Ehsan2019Transf} prove that majority of the different cases, the attacks are not transferable in image forensics particularly between different neural networks. 
	Therefore, additional study is required on the attacker’s side to know if and how the transferability can be improved, for example, by improving the attack strength or proposing the most powerful attack. We expect that by increasing the strength of attacks, adversarial attacks can be transferred to the target network even in mismatch networks (see Figure \ref{Boundary.Fig}). We expect that future work will concentrate more on improving attacks' strength to increase adversarial examples' transferability. Given the complexity of the decision margin learned by the CNNs, controlling the amount of distortion applied by the attacker on the image, is not an easy task. \\ Another direction for future research is related to randomization-based defense when attackers aware of the mechanism, such as feature selection and the architecture. Therefore, we expect that the use of different FC layers or different kernels in SVM can improve the security mechanism on the defender side.
	\begin{figure}[h!]
		\centering
		\includegraphics[width=0.5\columnwidth]{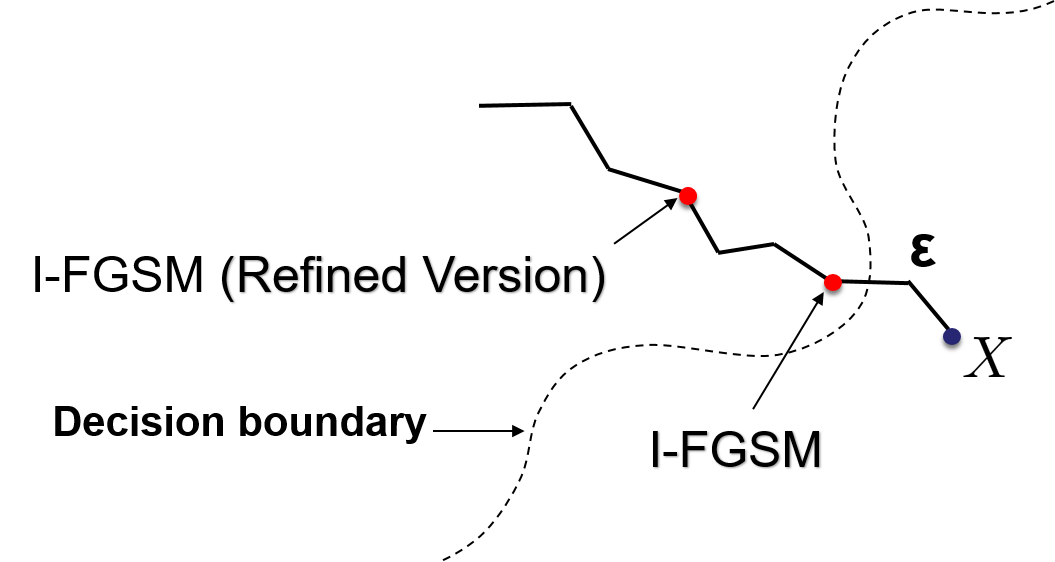}
		\caption{Increasing strength of the attack when adversarial example enters more inside the decision boundary}
		\label{Boundary.Fig}		
	\end{figure}

	\item \textbf{\textit{GAN-CF detection need more investigation}} In recent years with advancements in DL, several techniques have been extended to create fake image contents, e.g., such as GANs \cite{Goodfellow2014GAN,Kim2017CNNantiMedian}. GANs have extensively been utilized to generate synthetic images visually indistinguishable from real ones. Therefore we need to develop mechanisms to differentiate fake from realistic images. Most of the existing methods mainly focus on improving GAN image detection with high detection performance without consideration of anti-forensics. Anti-forensics are also recently utilized for these kinds of detection tools, aiming to attack forensic techniques where the attacker might apply GANs to train a generator to counterfeit forensic traces. \\
	To expose these detection method's weaknesses, an attacker can devise a generator anti-forensically to deceive most detection techniques without applying artifacts into the image. Further research in image forensics is needed to understand how GAN-CF can be detected with innovative approaches. 
	
	
	\item \textbf{\textit{Defense against backdoor attacks}} Another new class of attacks against DL architectures is backdoor attacks \cite{Tianyu2017}.. Backdoor attacks are in the category of causative attacks when training sets are poisoned by introducing a backdoors signal to a portion of the training files that cause misclassification of a payload during a testing time, which poses a new and realistic threat, particularly in image forensic applications. Although several works have been done, difficulties still exist due to the problems' complexity and forensic applications' wideness. \\
	Investigating solutions to enhance the security of ML techniques facing such attacks in image forensics plays an important role. For instance, randomized smoothing is one of the strategies newly considered against backdoor attacks, which was extended to verify robustness versus adversarial attacks but still have limited effectiveness. Another challenges is related to semantic and physical backdoor attacks, and researches are left far behind. Therefore, a better perception of those areas would be an essential step towards defeating the backdoor threat in practice.
	
	
	\item \textbf{\textit{Developing security class of ML}} Machine learning (ML) techniques have proved remarkable performance in numerous application fields, particularly in image forensics. Traditionally, most of ML methods are trained in a benign setting with identical statistical characteristics in training and testing. Whereas, this assumption usually in ML does not work against adversarial attacks, where some statistical features of the samples can be tampered with by an intelligent adversary. Therefore, one way is to design an ML model in an adversarial setting. The susceptibility of ML techniques in adversarial environments and corresponding countermeasures need to investigate more.  \\ Future research on ML security, in particular reference to image forensics, may involve the following aspects. The defense strategy facing adversarial attacks for DL techniques should include more studies. Many studies proved that DNNs are inherently fragile to subtle perturbations that affect output results. Although numerous defense approaches have recently been proposed so far to counter such attacks, we can not find a primary solution to these difficulties. Thus, generating a secure ML/DL under adversarial settings will likely represent a significant challenge for the years to come.

	\item \textbf{\textit{Deep learning interpretability}} Interpretability in deep learning-based methods is another hot issue, particularly in image forensic tools. Understanding the nature of the black-box generally is difficult and how the network makes a decision. To clarify more, in pattern recognition, DL can accurately classify cats from dogs, but we can not understand which specific features lead to this decision. This problem also happens in most forensic applications. Hence, tracking DL's reasoning would enable to improve most of the forensic networks in terms of security and robustness against adversarial attacks.

   \item \textbf{\textit{Multiple adversarial attacks to ML}} This problem happens in most ML/DL image forensic applications. The challenge considerably different when the ML system has to face multiple adversaries. Based on the synthesization of knowledge (SoK) in the survey, only a few papers have studied models with multiple adversaries. Thus, colluding attacks within Adversarial Image Forensics is a hot issue that can be studied further.
	%
	
	

%
	\EN{.} \\ 
	
\end{itemize}
\section*{Acknowledgements}
The first author thanks members of the Visual Information Processing and Protection (VIPP) group at the University of Siena, Italy for their suggestions.

\bibliographystyle{elsarticle-num}
\bibliography{mybibfile}

\end{document}